\documentclass[fleqn,usenatbib]{mnras}

\usepackage{mathptmx}

\usepackage[T1]{fontenc}

\DeclareRobustCommand{\VAN}[3]{#2}
\let\VANthebibliography\thebibliography
\def\thebibliography{\DeclareRobustCommand{\VAN}[3]{##3}\VANthebibliography}

\usepackage{graphicx}	
\usepackage{amsmath}	
\usepackage{amssymb}	
\usepackage{comment}
\def \LGalaxies{\texttt{L-Galaxies}\,}
\def \fm{\textit{fiducial model}}
\def \dm{\textit{delayed model}}

\usepackage{ulem}
\def \mbin{MBHBs}
\def \msun{\,\rm M_\odot}
\def \mBHS{MBHs}
\usepackage{color}
\definecolor{myorange}{rgb}{0.8, 0.3, 0.0}

\def\msun{\,\rm{M_\odot}}

\title [Hosts and signatures of massive binaries]{Unveiling the hosts of parsec-scale massive black hole binaries: morphology and electromagnetic signatures}
\author[Izquierdo-Villalba et al.]{David Izquierdo-Villalba,$^{1,2}$\thanks{E-mail: david.izquierdovillalba@unimib.it}
Alberto Sesana,$^{1,2}$
and Monica Colpi$^{1,2}$
\\
$^{1}$ Dipartimento di Fisica ``G. Occhialini'', Universit\`{a} degli Studi di Milano-Bicocca, Piazza della Scienza 3, I-20126 Milano, Italy\\
$^{2}$ INFN, Sezione di Milano-Bicocca, Piazza della Scienza 3, 20126 Milano, Italy\\
}

\date{Accepted XXX. Received YYY; in original form ZZZ}

\pubyear{2015}

\begin{document}
\label{firstpage}
\pagerange{\pageref{firstpage}--\pageref{lastpage}}
\maketitle

\begin{abstract}
Parsec-scale massive black hole binaries (\mbin{}) are expected to form in hierarchical models of structure formation. Even though different observational strategies have been designed to detect these systems, a theoretical study is a further guide for their search and identification.  In this work, we investigate the hosts properties and the electromagnetic signatures of massive black holes gravitationally bound on parsec-scales with primary mass $\rm {>}\,10^7\,M_{\odot}$. For that, we construct a full-sky lightcone by the use of the semi-analytical model \LGalaxies{} in which physically motivated prescriptions for the formation and evolution of \mbin{} have been included. Our predictions show that the large majority of the \mbin{} are placed either in spiral galaxies with a classical bulge structure or in elliptical galaxies. Besides, the scaling relations followed by \mbin{} are indistinguishable from the ones of single massive black holes. We find that the occupation fraction of parsec-scale  \mbin{}  reaches  up to ${\sim}\,50\%$ in galaxies with $\rm M_{stellar}\,{>}\,10^{11}\, M_{\odot}$ and drops below 10\% for $\rm M_{stellar}\,{<}\,10^{11}\, M_{\odot}$. Our model anticipates that the majority of parsec-scale \mbin{} are unequal mass systems and lie at $z\,{\sim}\,0.5$, with ${\sim}\,20$ objects per $\rm deg^2$ in the sky. However, most of these systems are inactive, and only $1\,{-}\,0.1$ objects per $\rm deg^2$ have an electromagnetic counterpart with a bolometric luminosity in excess of $10^{43}$ erg/s. Very luminous phases of parsec-scale \mbin{} are more common at $z\,{>}\,1$, but the number of binaries per $\rm deg^2$ is ${\lesssim}\,0.01$ at $\rm L_{\rm bol}\,{>}\,10^{45} \rm erg/s$.
\end{abstract}




\section{Introduction}
According to our current hierarchical structure formation paradigm, the assembly of galaxies takes place through major mergers, accretion of satellites onto larger galaxies, and accumulation of intergalactic gas funneled into dark matter filaments \citep{WhiteandRees1978,WhiteFrenk1991,Haehnelt1993,Kauffmann1999}. Several observations further  demonstrate that massive black holes (\mBHS{}, \rm $\rm {>}\,10^{6}\, M_{\odot}$) reside at the centres of today galaxies and they were the engine of quasars and active galactic nuclei at earlier cosmic times \citep{Genzel1987,Kormendy1988a,Dressler1988,Kormendy1992,Genzel1994,ODowd2002,HaringRix2004,Peterson2004,Vestergaard2006,Hopkins2007,Kormendy2013,Savorgnan2016,Shen2020}. By joining this information together, massive black hole binaries (\mbin{}) are expected to form  in the aftermath of galaxy collisions.\\

The evolution of \mbin{} have been explored in several theoretical studies. Our current framework is based on the pioneering work of \cite{Begelman1980} which argued that after the  merger of two galaxies, their central \mBHS{} are dragged towards the center of the remnant galaxy under the action of dynamical friction caused by background stars and gas forming a binary, i.e. a gravitational bound system on a scale of a parsec (pc) \citep{Chandrasekhar1943}. When reaching sub-parsec scales the MBHB hardens interacting with individual stars  \citep{Quinlan1997,Sesana2006,Vasiliev2014,Sesana2015} and/or torques extracted from a circumbinary gaseous disc \citep{Escala2004,Escala2005,Dotti2007,Cuadra2009,Franchini2021}. On milli-parsec scales  gravitational waves (GWs) becomes the main mechanism of extracting  energy and angular momentum from the system, eventually driving the two black holes to coalescence on  timescales between Myrs to several Gyrs \citep[see][for a review]{Colpi2014}.  During the GW driven phase, the MBHB is a loud source of GWs, whose  frequency ${<}\,\rm Hz,$ equal or higher than the Keplerian frequency,  depends on the separation between the two black holes and their masses. Current and future experiments aim at detecting the GWs generated by these systems. While the lowest frequencies ($10^{-9}\,{-}\,10^{-7}\,\rm Hz$) are generated by MBHBs above $\rm 10^8 \, M_{\odot}$ and are the main target of Pulsar Timing Array experiments \citep[PTA,][]{Sesana2004,Sesana2008,Sesana2009,Hobbs2010,Desvignes2016,Arzoumanian2015,Reardon2016,Bailes2016,Perera2019,Susobhanan2021}, the frequency range ${\sim}\,10^{-5} \, \rm Hz \,{-}\, 0.1\,\rm Hz$, covered by coleascing MBHs of $\rm 10^{4} \,{-}\, 10^{7}\, M_{\odot}$, will be probed by space-based interferometers such as ESA's Laser Interferometer Space Antenna \citep[LISA,][]{LISA2017} and the Chinese concepts TianQin and Taiji \citep{Luo2016,WenRui2017,Ruan2018}.\\

PTAs recently reported the detection of a clear common red signal in their data  \citep{Arzoumanian2020,Goncharov2021,Chen2021}. Although its origin is still unknown, it might be the signature of the long-sought-after GW astrophysical background from inspiraling  MBHBs with orbital periods of ${\sim}\,0.1\,{-}\,10$ years and masses above $10^{8}\msun$ located at $z\,{\lesssim}\,1$ \citep{Sesana2009,Sesana2012,Kelley2016,IzquierdoVillalba2021}. Albeit indirectly, this will inform on the presence of galaxies hosting parsec-scale MBHBs on their way to merge. PTA observations might also detect individual sources if sufficiently loud and nearby to emerge from the background \citep{Rosado2015,Kelley2018,Goldstein2019}.  Characterizing the properties of the galaxies hosting these parsec-scale MBHBs 
is compelling and theoretical investigations have been carried on by a number of authors \citep[see e.g.][]{Tanaka2012,Kelley2016}. In particular, \cite{MartinezPalafox2014}, using  N-body simulations and semi-empirical approaches to populate galaxies with \mbin{}, found that binary systems with $\rm M_{BH}\,{>}\,10^8\, M_{\odot}$ are hosted in  halos with masses $\rm {>}\,10^{12}\, M_{\odot}$ and have typically mass ratios  $q\,{>}\,0.1.$   Similar conclusions were reached by \cite{Tanaka2012} and \cite{Rosado2014} which reported that the most massive binaries emitting GW at ${\sim}\,\rm nHz$ frequencies should be hosted in dense environments and massive galaxies. The possibility of using signatures of recent galaxy interactions as a plausible indicator of the MBHB presence was recently studied in hydrodynamical simulations by \cite{Volonteri2020}. The results pointed out that large delays between the galaxy merger and the final MBHB coalescence is expected to wash out  distinctive morphological signatures indicating ongoing galaxy interaction.\\

Besides galaxy properties, direct electromagnetic signatures resulting from gas accretion onto the two holes can be used as a powerful tool to detect the presence of \mbin{} \citep[see the recent reviews of][]{Derosa2019,Bogdanovic2021}. In fact, given that these systems form in galaxy mergers, strong gas inflows towards the center of the remnant galaxy are expected \citep{DiMatteo2005,Springel2005,Hopkins2009c}. During such process part of the gas can reach the vicinity of the \mBHS{}, causing the formation of a circumbinary accretion disc around the MBHB, inducing observational signatures in the form of electromagnetic signals \citep{Escala2005,Cuadra2009}. To shed light on the properties of these signatures, several work have been done. By performing a simulation of the prototypical circumbinary disc around binaries with $\rm M_{BH}\,{>}\,10^8\, M_{\odot}$, \cite{Sesana2012} studied the possible counterparts in the X-ray domain. Interestingly, the authors found that double broad $\rm K\alpha$ iron lines could be a good indicator for the existence of massive binaries. A similar study was performed by \cite{Tanaka2009} but using an analytical approach. The authors reported that active binaries with $\rm M_{BH}\,{>}\,10^8\, M_{\odot}$ might be unusually low X-ray sources when compared to single active \mBHS{} with the same total mass. Another alternative way to detect the presence of a \mbin{} is through a variability analysis. As shown in the theoretical work of, e.g., \cite{Kelley2019}, the binary period can imprint variations in the luminosity emitted by the \mbin{}. The authors reported that the Doppler boosting due to large orbital velocities or hydrodynamic variability caused by the interaction between accretion discs and the \mBHS{} could lead to a distinctive variability signature. 
In quantitative terms, \cite{Kelley2019} predicted that hundreds of \mbin{} could be detected by photometric variability studies with 5 years of Large Synoptic Survey Telescope observations \citep{LSST2019} (see also \citealt{Charisi2022}). On the observational side, while dual active galactic nuclei (AGN) on kpc scales have been observed at different wavelengths \citep{Rodriguez2006,Koss2012,Comerford2012}, up to date there are no confirmed AGNs associated to gravitational bound binaries. This is principally caused by the fact that at the early evolutionary stages of a galaxy merger, the two \mBHS{} can still be spatially resolved. However, when the two \mBHS{} reach the galactic center and get gravitational bound, their angular separation in the sky (${<}\,0.01\, \rm arsec$) is beyond the resolution capabilities of current instrumentation. To overcame this limitation, searching techniques rely on indirect measurements such as periodic lightcurves of AGNs. While hundred of these periodic systems have been suggested to be the manifestation of \mbin{} \citep[see e.g.][]{Valtonen2008,Graham2015,Charisi2016,Liu2016,Liu2019,Liao2021,Witt2021}, it is still unclear what is the number of false positives due to AGN stochastic variability \citep[see e.g.][]{Vaughan2016}. Besides periodicity signatures, the presence of Doppler-shifting of AGN broad emission lines due to the binary's orbital motion has been proposed and searched for a distinctive feature of sub-parsec \mbin{} \citep[see e.g.][]{Bogdanovic2009,Tsalmantza2011,Montuori2011, Eracleous2012,Shen2013}. However, one of the main limitations of these studies concerns the difficulty of tracing with high accuracy the variability pattern of the broad lines, fundamental to study the existence of orbital motion related to parsec-scale \mbin{} \citep[][]{Eracleous2012}.\\

The purpose of this paper is characterize the morphological properties of the galaxies hosting $\rm {>}\,10^7\, M_{\odot}$ parsec-scale MBHBs and their electromagnetic signatures in further depth, by using a simulated lightcone detailed enough to account for the cosmological evolution of galaxies and their embedded single and dual MBHs, over a wide range of masses and binary separations. With the mock universe, we aim at providing first quantitative estimates on the occupation fraction of parsec-scale MBHBs, the number of these objects per $\rm deg^2$, and how many of them can be seen in an active phase. For this purpose, we make use of the \LGalaxies{} semi-analytical model \citep[SAM,][]{Henriques2015} applied on the \texttt{Millennium} merger trees \citep{Springel2005}. The version of the SAM used in this work is the one presented in \cite{IzquierdoVillalba2021} which includes physically-motivated prescriptions to treat the physical processes driving MBHBs. Indeed, the model has been tested to be consistent with the PTA limits on the GW background. On top of this, the version of \cite{IzquierdoVillalba2021} has been modified to include the procedure presented in \cite{IzquierdoVillalba2019LC} which let us transform the simulated boxes provided by \LGalaxies{} into a lightcone.\\

The paper is organised as follows: In Section~\ref{sec:SAM_DESCRIPTIONS} we describe the main characteristics of the \texttt{Millennium} simulation and \LGalaxies{}, and  the procedure used to create a lightcone in which the physics of galaxies, MBHs and the dynamics of MBHBs have been taken into account. In Section~\ref{sec:GrowthModels} we explore two different physically motivated models for tracking the mass growth of MBHs and MBHBs.  In Section~\ref{sec:HostsofMBHBs} and \ref{sec:ElectromagneticCounterpart} we present the properties of the galaxies hosting \mbin{} and explore the electromagnetic counterparts of these binaries, respectively. Finally, in Section~\ref{sec:Conclusions} we summarize our main findings. A Lambda Cold Dark Matter $(\Lambda$CDM) cosmology with parameters $\Omega_{\rm m} \,{=}\,0.315$, $\Omega_{\rm \Lambda}\,{=}\,0.685$, $\Omega_{\rm b}\,{=}\,0.045$, $\sigma_{8}\,{=}\,0.9$ and $\rm H_0\,{=}\,67.3\, \rm km\,s^{-1}\,Mpc^{-1}$ is adopted throughout the paper \citep{PlanckCollaboration2014}.

\section{A LIGHTCONE FOR THE STUDY OF MASSIVE BLACK HOLE BINARIES} \label{sec:SAM_DESCRIPTIONS}

In this section we present the galaxy formation model used to generate a lightcone tailored for the study of pasrec-scale MBHBs. Specifically, we make use of the \LGalaxies{} semi-analytical model, a state-of-the-art SAM able to reproduce many different observational scaling relations such as the stellar mass function, the cosmic star formation rate density, galaxy colors and the fraction of passive galaxies (we refer to \citealt{Guo2011} and \citealt{Henriques2015} for further details). In particular, the version of \LGalaxies{} used in this work is the one presented in \cite{Henriques2015} with the modifications of \cite{IzquierdoVillalba2019,IzquierdoVillalba2020,IzquierdoVillalba2021}. These changes were included to improve the galaxy morphology, extend the physics of MBHs and introduce the formation and evolution of \mbin{}. In the following, we summarize the main features of the model, and we refer the reader to the papers cited above for a detailed description of the baryonic physics included.

\subsection{Generating a galaxy population: Dark matter and baryons} \label{sec:MillLGal}
The backbone of \LGalaxies{} are the subhalo (hereafter just halo) merger trees of the \texttt{Millennium} \citep[MS,][]{Springel2005} and \texttt{Millennium-II} \citep[MSII,][]{Boylan-Kolchin2006} dark matter (DM) only simulations. In this work we use the MS, whose mass resolution allows us to study the evolution of \mBHS{} and \mbin{} hosted in galaxies with stellar mass $\rm {>}\,10^9\,M_{\odot}$. In brief, the MS follows the cosmological evolution of $2160^3$ DM particles with mass $8.6 \times 10^8\, \mathrm {M_{\odot}}/h$ within a periodic cube of $500^3\,{\rm Mpc^3}/h^{3}$ comoving volume. Halos were extracted using the \texttt{SUBFIND} algorithm and arranged according to their evolutionary path in the so-called merger trees \citep{Springel2001,Springel2005}. These trees contain the information of all MS halos at 63 different epochs or \textit{snapshots}. To improve the tracing of the baryonic physics, \LGalaxies{} does an internal time interpolation between two consecutive snapshots with approximately $\rm{\sim}\,5{-}20 \,Myr$ of time resolution depending on redshift. Finally, by using the methodology of \cite{AnguloandWhite2010}, \LGalaxies{} re-scales the original cosmological parameters of the MS to match the ones provided by the Planck first-year data release \citep{PlanckCollaboration2014}.\\

Regarding the baryonic component, \LGalaxies{} assumes that the birth of a galaxy starts through the infall of baryonic matter onto the newly formed DM halo \citep{WhiteFrenk1991}. This process is modeled by associating an amount of matter to each halo according to the cosmic mean baryon fraction. During the infalling, the baryonic component is expected to shock-heat and form a diffuse, pristine, spherical, and quasi-static hot gas atmosphere with an extension equal to the halo \textit{virial radius} ($R_{200c}$). Part of this hot gas is then allowed to cool down and migrate towards the DM halo center \citep{WhiteandRees1978}. The rate at which this process takes place is determined by the cooling functions of \cite{SutherlandDopita1993} and the amount of hot gas enclosed within the halo \textit{cooling radius} ($r_{\rm cool}$), defined as the radius at which the cooling time matches the halo dynamical time. This implies the presence of two different cooling regimes: the \textit{rapid infall} ($\rm r_{cool}\,{>}\,R_{200c}$) which provokes  the condensation of the whole hot atmosphere, and the slower \textit{cooling flow regime} ($\rm r_{cool}\,{<}\,R_{200c}$), in which only a fraction of the hot gas is allowed to cool down. This new cold material settles into a disc-like structure whose specific angular momentum is inherited from the host DM halo \citep[see][]{Guo2011}. Because of this fuel of cold gas, the protogalaxy is capable to assemble a stellar disc component through star formation processes, on a time scale given by the cold gas disc dynamical time. As a result of the star formation, massive and short-lived stars explode as supernovae (SNe) injecting energy into the cold gas disc (\textit{SNe feedback}), causing the re-heat of a fraction of it and expelling some of the hot gas beyond the halo virial radius. At later times, this ejected gas can be reincorporated, fuelling new star formation events. Thanks to all these processes, the stellar disc can be prone to non-asymmetric instabilities (or just disc instabilities) which ultimately lead to the formation of a central ellipsoidal component, called \textit{pseudobulge}. Besides SNe feedback, \LGalaxies{} introduces the feedback from the central MBHs as an additional mechanism to regulate the star formation in massive galaxies. This is the so-called \textit{radio-mode feedback} which is caused by the continuous gas accretion onto the MBH ($\dot{\rm M}_{\rm BH}$) from the galaxy hot gas atmosphere. This gas accumulation, which is typically orders-of-magnitude below the Eddington limit, is determined as \citep{Henriques2015}:
\begin{equation}\label{eq:Radio_mode}
\dot{\rm M}_{\rm  BH} \rm \,{=}\, \mathit{k}_{AGN} \left( \frac{M_{hot}}{10^{11}M_{\odot}} \right) \left( \frac{M_{BH}}{10^{8}M_{\odot}}\right),
\end{equation}
where $\rm M_{hot}$ is the total mass of hot gas surrounding the galaxy and $\rm \mathit{k}_{AGN}$ is a free parameter set to $\rm 3\,{\times}\,10^{-4} M_{\odot}/yr$ to reproduce the turnover at the massive end of the galaxy stellar mass function. During this low-accretion process, the MBH injects energy into the nearby medium, reducing or even stopping the cooling flow towards the DM halo center.\\

In addition to internal processes, the hierarchical growth of the DM halos shapes the galaxy properties. Galactic encounters in \LGalaxies{} are driven by the merger of the parent DM halos. The time-scale of these processes is given by the dynamical friction experienced by the merging galaxies, accounted by using the \cite{BinneyTremine1987} formalism. According to the baryonic mass ratio of the two interacting galaxies ($\rm m_R$), \LGalaxies{} distinguishes between \textit{major} and \textit{minor} interactions. On one hand, major mergers completely destroy the discs of the two galaxies, giving rise to a pure spheroidal remnant (\textit{elliptical}) which undergoes a \textit{collisional starburst}. On the other hand, during minor interactions the disc of the larger galaxy survives and experiences a burst of star formation, while its bulge integrates the entire stellar mass of the satellite (forming a \textit{classical bulge}). In addition, the model used in this work includes the prescription of \textit{smooth accretion} in order to deal with the physics of extreme minor mergers \citep{IzquierdoVillalba2019}, important to obtain the observed morphology of dwarf galaxies ($\rm M_{stellar}\,{\leq}\,10^9 \, M_{\odot}$). During these events it is expected that the stellar remnant of the satellite (comprehensive of the bulge and disc) gets diluted  inside the disc of the central galaxy before being able to reach the nucleus, thus, losing the possibility of make the bulge of the primary galaxy to grow.


\subsection{The population of massive black holes and massive black hole binaries} \label{sec:MBH_and_MBHBHs_Model}

Each halo is seeded with a MBH of $\rm 10^{4} M_{\odot}$\footnote{The large seed mass chosen in this work is motivated by the dark matter resolution of the \texttt{Millennium} simulation. The vast majority of the newly formed halos have masses $\rm {\sim}\,10^{10}\, M_{\odot}$. This implies that the SAM can not access the assembly of the galaxy and its central MBH before their parent dark matter halo is resolved. Therefore, to account for this unresolved evolution we decided to place a relatively massive MBH seed in each galaxy.} and random spin in the range $0\,{<}\,\chi\,{<}\,0.998$ (further improvements on the seeding paradigm will be done by including the model of \citealt{Spinoso2022}). Once the black hole seed is placed in its host galaxy, it can grow trough three different channels: \textit{cold gas accretion}, \textit{hot gas accretion}, and \textit{mergers} with other black holes. Specifically, the first channel is the main driver of the black hole growth and it is triggered by both galaxy mergers  and disc instability events. After a galaxy merger the model assumes that the nuclear black hole accretes a fraction of the galaxy cold gas given by:
\begin{equation}\label{eq:QuasarMode_Merger}
\rm   \Delta {M}_{BH}^{gas} \,{=}\,\mathit{f}_{BH}^{merger} (1+\mathit{z}_{merger})^{5/2} \frac{m_{R}}{1 + (V_{BH}/V_{200})^2}\, M_{\rm gas},
\end{equation}
where $\rm m_{R}\,{\leq}\,1$ is the baryonic ratio of the two merging galaxies, $z_{\rm merger}$ the redshift of the galaxy merger, $\rm V_{200}$ the virial velocity of the host DM halo, $\rm M_{\rm gas}$ the cold gas mass of the galaxy and $\rm V_{BH}$, $\rm \, \mathit{f}_{BH}^{\rm merger}$ two adjustable parameters set to $\rm 280 \, km/s$ and $0.014$, respectively. If a disc instability takes place in a galaxy, the black hole accretes an amount of cold gas proportional to the mass of stars that trigger the stellar instability, $\rm \Delta M_{\rm stars}^{DI}$:
\begin{equation}\label{eq:QuasarMode_DI}
\rm    \Delta {M}_{BH}^{gas} \,{=}\, \mathit{f}_{BH}^{DI} (1+\mathit{z}_{DI})^{5/2} \frac{\Delta M_{stars}^{DI}}{{1 + (V_{BH}/V_{200})^2}},
\end{equation}
where $\rm \mathit{z}_{DI}$ is the redshift in which the disc instability takes place, and $\rm \mathit{f}_{BH}^{DI}$ is a free parameter that takes into account the gas accretion efficiency, set to $0.0014$. All these adjustable parameters  have been tuned to give the best agreement between the observations and model predictions for the $z\,{=}\,0$ black hole-bulge correlation.\\

After a galaxy merger or a disc instability, the cold gas available for accretion ($\rm \Delta M_{BH}^{gas}$) is assumed to settle in a reservoir around the black hole, which is progressively consumed according to a two phases model, described in Section~\ref{sec:GrowthModels}. During any of the MBH growth events, \LGalaxies follows the evolution of the black hole spin modulus in a self-consistent way. During gas accretion events, the model links the number of accretion events that spin-up or spin-down the MBH with the degree of coherent motion in the bulge \citep[following][]{Sesana2014}. In particular, we assume that disc instabilities (mergers) increase (decrease) the coherence of the bulge kinematics. On the other hand, after a MBH coalescence the final spin is determined by the expression of \cite{BarausseANDRezzolla2009}.\\

\begin{figure}
\centering  
\includegraphics[width=1.0\columnwidth]{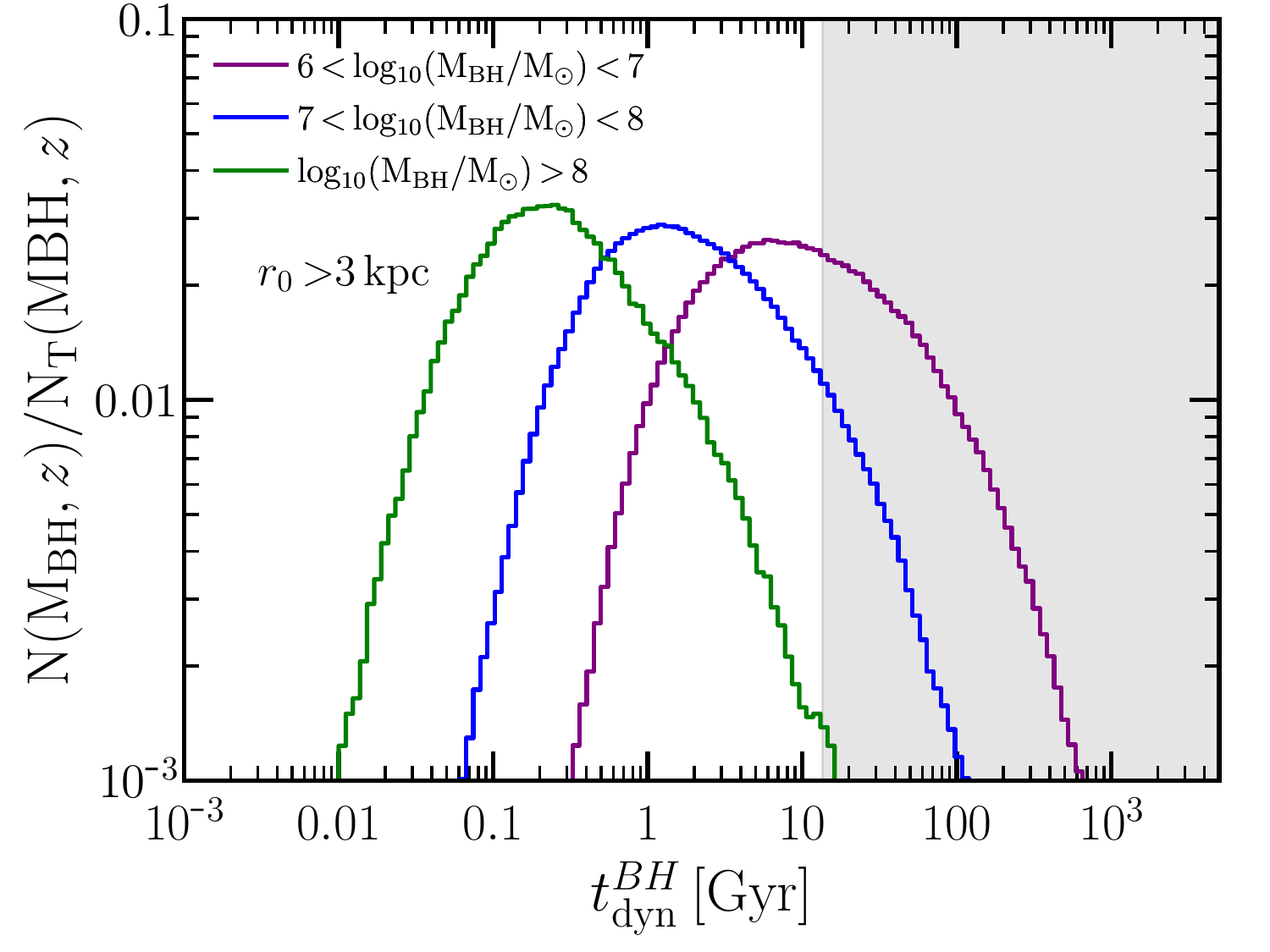}
\caption[]{Distribution of the dynamical friction time, $t^{\rm BH}_{\rm dyn}$, predicted by \LGalaxies{} for MBHs inhabiting satellite galaxies merging into larger halos. The MBHs  presented corresponds to the ones with $r_0\,{>}\,3\, \rm kpc$ and masses in the range  $\rm 6\,{<}\,log_{10}(M_{BH,1}/M_{\odot})\,{<}\,7$ (purple) $\rm 7\,{<}\,log_{10}(M_{BH,1}/M_{\odot})\,{<}\,8$ (blue), and $\rm log_{10}(M_{BH,1}/M_{\odot})\,{>}\,9$ (green). The grey shaded area highlights the times longer than the age of the Universe (${\sim}\,13\, \rm Gyr$)}
\label{fig:Dynamical_friction_time_Scale}
\end{figure}

On top of the evolution of single MBHs, \LGalaxies{} deals with the formation and dynamical evolution of MBHBs \citep{IzquierdoVillalba2021}. Following the picture of \cite{Begelman1980}, we divide the evolution of these systems in three different stages: \textit{pairing, hardening} and \textit{gravitational wave} phase. The first one starts after a galaxy merger. During this phase, dynamical friction caused by the stars of the remnant galaxy reduces the initial separation ($\rm {\sim}\,kpc$) between the nuclear and the satellite MBH, sinking the latter towards the galactic center. To determine the time needed for the satellite MBH to reach the nuclear part of the galaxy, $t_{\rm dyn}^{\rm BH}$, \LGalaxies{} uses the expression \citep{BinneyTremaine2008}:
\begin{equation}  \label{eq:DynamicalFriction}
    t_{\rm dyn}^{\rm BH} \,{=} \, 19 \, f(\varepsilon)  \left( \frac{r_0}{5 \, \rm kpc} \right)^2 \left( \frac{\sigma}{200 \,\rm km/s}\right) \left( \frac{10^8 \, \rm M_{ \odot}}{\rm M_{BH}} \right) \, \frac{1}{\Lambda}\, \rm [Gyr] , 
\end{equation}
where $f(\varepsilon)\,{=}\,\varepsilon^{0.78}$ is a function with depends on the orbital circularity of the MBH $\varepsilon$ \citep{Lacey1993}, $r_0$ is the initial position of the black hole deposited by the satellite galaxy after the merger, $\sigma$ is the velocity dispersion of the remnant galaxy, $\rm M_{BH}$ is the mass of the satellite (lighter) black hole and $\rm \Lambda\,{=}\,\ln(1 + M_{stellar}/M_{BH})$ is the Coulomb logarithm  \citep{MoWhite2010}. We refer the reader to \cite{IzquierdoVillalba2021} for further information about the calculations of these quantities.
The specific value of $t_{\rm dyn}^{\rm BH}$ strongly correlates with $r_0$ and $\rm M_{BH}$. To guide the reader, in Fig.~\ref{fig:Dynamical_friction_time_Scale} we present the $t_{\rm dyn}^{\rm BH}$ distribution predicted by \LGalaxies{} at three different bins of MBH masses for cases with $r_0\,{>}\,3\,\rm kpc$ (notice that no redshift cut has been imposed). As we can see, MBHs with masses between $\rm 10^6\,{-}\,M_{BH}\,{<}\,10^7 \, M_{\odot}$ experience dynamical friction phases of $t_{\rm dyn}^{\rm BH}\,{\sim}\,1\,{-}10\, \rm Gyr$. On the contrary, MBHs of $\rm M_{BH}\,{>}\,10^8 \, M_{\odot}$ undergo faster dynamical phases of $t_{\rm dyn}^{\rm BH}\,{\sim}\,0.1\,{-}1\, \rm Gyr$ \citep[see Figure 3 of][for a dependence with the stellar mass]{IzquierdoVillalba2021}.\\

\begin{figure*}
\centering  
\includegraphics[width=2.0\columnwidth]{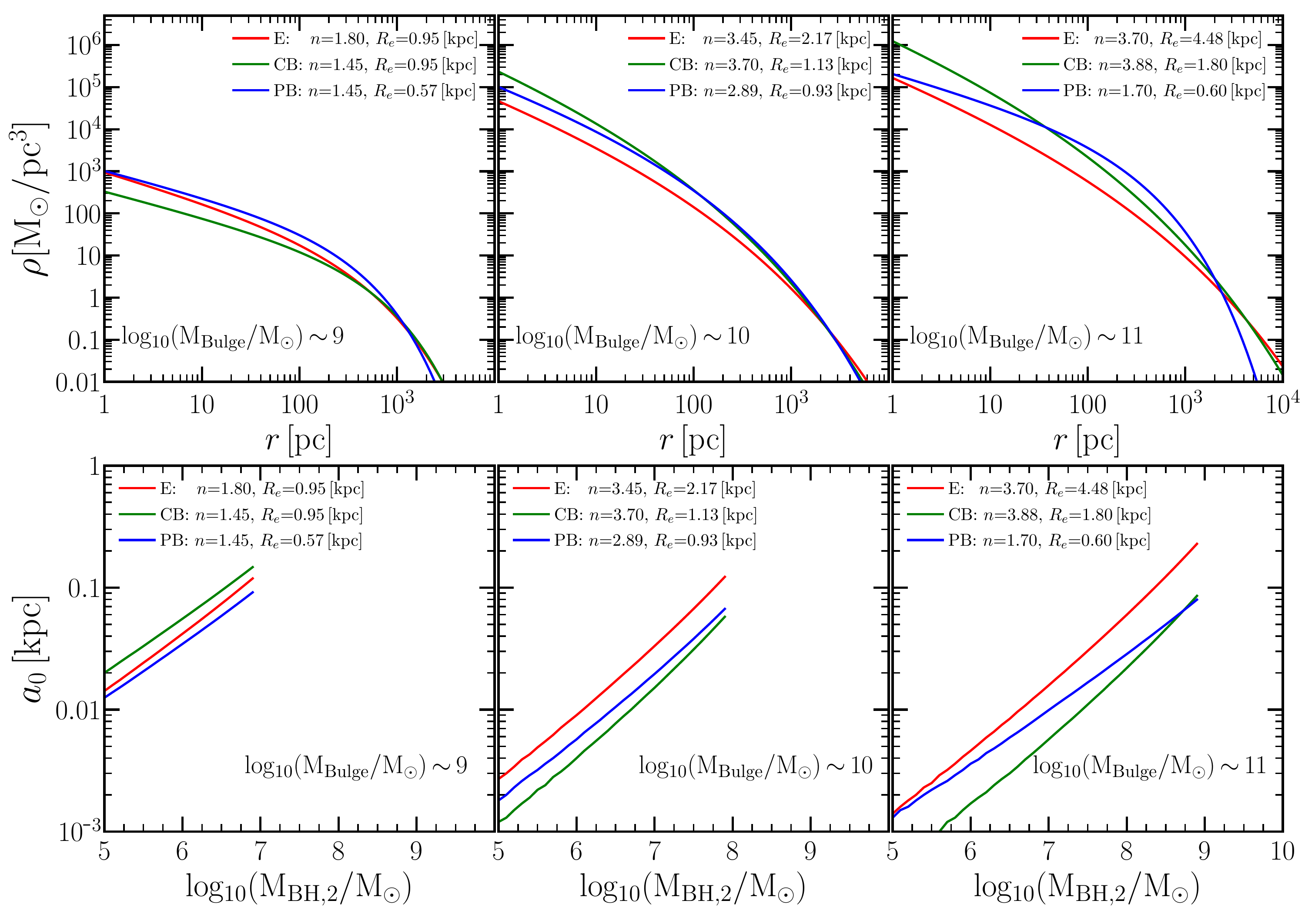}
\caption[]{\textbf{Upper panel}: Bulge mass density profile, $\rho(r)$. Left, middle and right panels show the results for a bulge mass of $\rm 10^9\,M_{\odot}$, $\rm 10^{10}\,M_{\odot}$ and $\rm 10^{11}\,M_{\odot}$, respectively. Red, green and blue lines represent the predictions for elliptical galaxies (E), classical bulges (CB) and pseudobulges (PB), respectively. The median structural properties ($n$, Sérsic index and $R_e$ bulge effective radius) of the bulge type in each bin of mass have been extracted form \protect{\cite{Gadotti2009}}. \textbf{Lower panel}: The initial separation of the binary at the beginning of the hardening phase, $a_0$, as a function of the mass of the secondary black hole, $\rm M_{BH,2}$. Line colors represent the same as in the upper panel. Motivated by the bulge-black hole mass correlation, in each panel we have only presented the results for $\rm M_{BH,2}\,{<}\, 0.01{\times}M_{Bulge}$.}
\label{fig:rho_a0_E_CB_PB}
\end{figure*}

Once the satellite MBH reaches the galaxy nucleus (i.e when the dynamical friction phase set by $t_{\rm dyn}^{\rm BH}$ is terminated), it binds with the central MBH ($\rm {\sim}\,pc$ separation) and the so-called \textit{hardening phase} begins. From hereafter, we assume that at this moment the two MBHs form a binary. We refer as \textit{primary black hole} (with mass $\rm M_{BH,1}$) the most massive black hole in the system whereas the less massive one is tagged as \textit{secondary black hole} (with mass $\rm M_{BH,2}$). The initial eccentricity of the binary, $e_0$, is selected randomly between $[0\,{-}\,1]$ while the initial separation, $a_0$, is set to the scale in which $\rm M_{Bulge}({<}\mathit{a_0})\,{=}\,2\,M_{BH,2}$, where $\rm M_{Bulge}({<}\mathit{a_0})$ corresponds to the mass in stars of the hosting bulge within $a_0$. To determine $a_0$, \LGalaxies{} assumes that the bulge mass density profile, $\rho(r)$, follows a Sérsic model \citep{Sersic1968}:
\begin{equation}\label{eq:Sersic}
 \rho (\mathit{r}) = \rho_0 \left( \frac{\mathit{r}}{R_e}\right)^{-p} \mathit{e}^{-b\left(\frac{\mathit{r}}{R_e}\right)^{1/n}},
\end{equation}
where $R_e$ is the bulge effective radius, $\rho_0$ is the central bulge density and $n$ its Sérsic index. Finally, $p$ and $b$ are two different quantities that depend on the Sérsic index of the bulge:  $p \,{=} \, 1 - 0.6097 n^{-1} + 0.05563 n^{-2}$ and $b\,{=}\, 2n - 0.33 + 0.009876n^{-1}$ \citep{Marquez2000}. Following \cite{Terzic2005} we can transform the density of Eq.~\ref{eq:Sersic} into mass and determine $a_0$ by solving:
\begin{equation}\label{eq:a0}
    \gamma \left(n(3-p),b\frac{a_0}{R_e}\right) \,{=}\,\frac{\rm M_{BH,2}}{2\pi \,\rho_0 \, R_e^3\, n\, b^{n(p-3)}},
\end{equation}
where $\gamma$ is the incomplete gamma function. Given that Eq.~\ref{eq:a0} needs to be solved numerically, in Fig.~\ref{fig:rho_a0_E_CB_PB} we present the typical values of $a_0$ and $\rho(r)$ for three different bulge masses, $\rm M_{Bulge}$. For each mass, we have explored three different parameter combinations to mimic the properties of three different types of galactic bulges: elliptical galaxies, classical bulges, and pseudobulges. Instead of employing arbitrary values, we have used as a reference the median parameters estimated by \cite{Gadotti2009} who explored the structural properties of $z\,{=}\,0$ bulges by using images of \textit{Sloan Digital Sky Survey}. As we can see, due to the different values of $n$ and $R_{e}$, pseudobulges, classical bulges and elliptical structures display a different behavior in $\rho(r)$ and $a_0$. For instance, massive ellipticals display the largest values of $a_0$ given that they are less centrally concentrated, more extended, and have lower densities. Concerning the typical values of $a_0$ for $\rm M_{BH,2}\,{\sim}\,10^5\,{-}10^6\, M_{\odot}$ we get $a_0\,{\lesssim}\,\rm 30\,pc$ whereas for $\rm M_{BH,2}\,{\sim}\,10^7\,{-}10^9\, M_{\odot}$ the values can increase up to $\rm {\sim}\,300\,pc$. \LGalaxies{} has the advantage of computing self-consistently the redshift evolution of the bulge mass, effective radius, and type (classical bulge, elliptical, and pseudobulge, see \citealt{Guo2011,IzquierdoVillalba2019}). Thus, we do not have to do any assumptions to get the values of $\rho_{0}$ and $R_e$. However, it does not provide information about the bulge Sérsic index. To attach a Sérsic value to each galaxy, we follow the methodology presented in \cite{IzquierdoVillalba2021}. In brief, based on the observational data of \cite{Gadotti2009} it is computed and fitted the Sérsic index distribution of $z\,{=}\,0$ pseudobulges, classical bulges, and ellipticals. Based on these distributions and the bulge type, a random Sérsic index is assigned to each simulated bulge. In brief, observational data of \cite{Gadotti2009} are fitted to infer the Sérsic index distribution of $z\,{=}\,0$ pseudobulges, classical bulges, and ellipticals. Based on these distributions and the bulge type, a random Sérsic index is assigned to each simulated bulge. We stress that using the Sérsic index of local galaxies for high-$z$ is a simplification of the model.\\


Once the hardening phase is started, we assume that the binary system can reduce its semi-major axis, $a_{\rm BH}$, through two different processes. If the system is surrounded by a gas reservoir with a mass larger than the mass of the binary, the interaction with a massive gaseous circumbinary disc is the principal mechanism of shrinking the binary separation (\textit{gas hardening}). In this case, \LGalaxies{} tracks the evolution of $a_{\rm BH}$ following the formalism of \cite{Dotti2015} and \cite{Bonetti2018a}:
\begin{equation}\label{eq:BH_separation_GW_inspiral_gas}
\begin {split}
\left( \frac{d a_{\rm BH}}{dt}\right)_{\rm Gas} \,{=}\, - \frac{2\dot{\mathrm{M}}_{\rm Bin}}{\mu} \sqrt{\frac{\delta}{1-e_{\rm BH}^2}} \, a_{\rm BH},
\end{split}
\end{equation} 
where $G$ is the gravitational constant, $c$ the light speed, $\delta\,{=}\,(1+q)(1\,{+}\,e_{\rm BH})$, $q\,{=}\,\rm M_{BH,2}/M_{BH,1}$ the mass ratio of the binary, $\dot{\mathrm{M}}_{\rm Bin}$ the sum of the accretion rate of both MBHs in the system (see Section~\ref{sec:GrowthModels}) and $\mu$ is the reduced mass of the binary. During this process, the eccentricity of the binary, $e_{\rm BH}$, is kept constant to a value of $0.6$ \citep{Roedig2011}. On the other hand, if the binary is placed in a gas poor environment the shrinking of $a_{\rm BH}$ is caused by three-body interactions with single stars \citep[\textit{stellar hardening},][]{Quinlan1997}. In this case, the values of $a_{\rm BH}$ and $e_{\rm BH}$ are tracked by integrating numerically the equations presented in \cite{Sesana2015}:
\begin{equation}\label{eq:BH_separation_GW_inspiral}
\begin {split}
\left( \frac{d a_{\rm BH}}{dt}\right)_{\rm Stars} \,{=}\, -\frac{G H \rho_{\rm inf}}{\sigma_{\rm inf}} a_{\rm BH}^2,
\end{split}
\end{equation}

\begin{equation} \label{eq:eccEvol_Stars}
\begin{split}
\left( \frac{d e_{\rm BH}}{dt}\right)_{\rm Stars} \,{=}\, \frac{G \rho_{\rm inf} HK}{\sigma_{\rm inf}}  a_{\rm BH},
\end{split}
\end{equation}
where $H\,{\approx}\,15\,{-}\,20$ and $K$ are respectively the hardening and eccentricity growth rate extracted from the tabulated values of \cite{Sesana2006}. The values of $\rho_{\rm inf}$ and $\sigma_{\rm inf}$ correspond respectively to the density and velocity dispersion of stars at the MBHB sphere influence, computed according to a Sérsic model. \\

Once the binary reaches sub-parsec scales, it enters in the \textit{gravitational wave inspiral} phase which drives the system to the final coalescence. In this phase, \LGalaxies{} follows the evolution of $a_{\rm BH}$ and $e_{\rm BH}$ by using the model of \cite{Sesana2015}:
\begin{equation}\label{eq:BH_separation_GW_inspiral_GW}
\begin {split}
\left( \frac{d a_{\rm BH}}{dt}\right) _{\rm GW} \,{=}\, \frac{64G^3 (\mathrm{M}_{\rm BH_1}\,{+}\,\mathrm{M}_{\rm BH_2})^3 F(e_{\rm BH})}{5c^5(1+q)^2 a_{\rm BH}^3},
\end{split}
\end{equation}

\begin{equation} \label{eq:eccEvol_GW}
\begin{split}
\left( \frac{d e_{\rm BH}}{dt}\right) _{\rm GW} \,{=}\, - \frac{304}{15} \frac{G^3 q (M_{\rm BH_1} + M_{\rm BH_2})^3}{c^5(1+q)^2 a_{\rm BH}^4 (1-e_{\rm BH}^2)^{5/2}} \left( e_{\rm BH} + \frac{121}{304}e_{\rm BH}^3\right),
\end{split}
\end{equation}
where $F(e_{\rm BH})$ is a function which depends on the binary eccentricity \citep{PetersAndMathews1963}:
\begin{equation}
F(e_{\rm BH}) \,{=}\, (1-e_{\rm BH})^{-7/2}\left[ 1 + \left( \frac{73}{24} \right)e_{\rm BH}^2  + \left( \frac{37}{96} \right)e_{\rm BH}^4 \right].
\end{equation}

If the lifetime of an MBHB is long enough, a third black hole impinging on the galaxy can reach the galaxy nucleus and interact with the MBHB system. If this happens, the interaction between the three MBHs can lead to multiple scenarios. In order to address the outcome of these events, \LGalaxies treats the triple black hole interaction by using the model of \cite{Bonetti2018ModelGrid}. In particular, tabulated values from \cite{Bonetti2018ModelGrid} are used to select those triple interactions which lead to the merger of a pair of MBHs and those causing the ejection of the lightest MBH from the system. In this latter case, the separation of the leftover MBHB is computed following \cite{Volonteri2003} and the resulting $e_{\rm BH}$ is select as a random value between $[0\,{-}\,1]$. To guide the reader, the typical result of the majority of these interactions is that the primary MBH does not change while the new secondary MBH is the most massive object between the secondary MBH of the already existing MBHB and the MBH that finished its dynamical friction phase.\\

\subsection{Lightcone construction}

One of the main objectives of this work is generating a realistic population of galaxies, \mBHS{} and \mbin{} that can be compared directly with observations, i.e taking into account similar selection effects, flux limiting observations, line-of-sight contamination or cosmic variance effects. Therefore, we need to transform the discrete number of comoving boxes provided by \texttt{Millennium} into a lightcone, i.e a mock Universe in which only galaxies, whose light has just enough time to reach the observer, are included. For that, we implemented in our version of \LGalaxies{} the method presented in \cite{IzquierdoVillalba2019LC}. In the following we briefly describe the procedure. \\

\begin{figure*}
\centering  
\includegraphics[width=1.75\columnwidth]{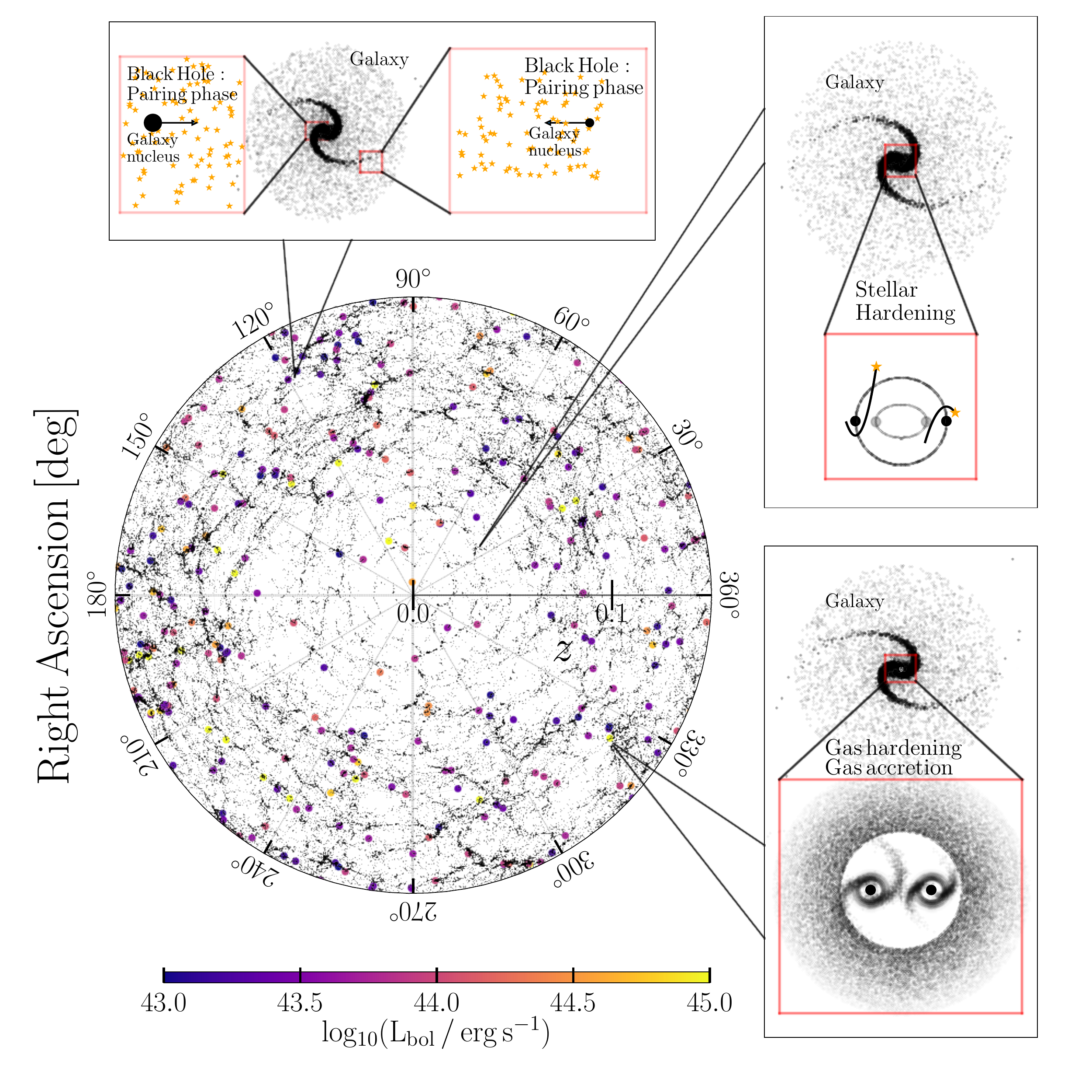}
\caption[]{The central plot displays the right ascension-redshift distribution of galaxies at $z\,{<}\,0.15$ with $\rm M_{stellar}\,{>}\,10^{10}\, M_{\odot}$ in a thin angular slice ($1\, \rm deg$) inside our lightcone. The colored points highlights the position of \mbin{} with bolometric luminosity $\rm L_{bol}\,{>}\,10^{43}\, \rm erg/s$ (the color encodes the value of $\rm L_{bol}$). The cartoons in the upper and left panels illustrate the different stages of binary evolution found in our lightcone. The upper panel represents the pairing phase, where the right panels display the stellar (top) and gas (bottom) hardening phase.}
\label{fig:LightCone}
\end{figure*}

The side-length of the \texttt{Millennium} simulation only allow to construct a lightcone with  $z\,{\sim}\,0.1$ depth. To extend this limit and reach larger redshifts, we take advantage of the periodic boundary conditions of \texttt{Millennium} and we replicate the box 15 times in each Cartesian coordinate. In this way we can generate a lightcone with a depth of $z\,{\sim}\,4$, regardless of the line-of-sight. Once complete, we place the observer at the origin of the first replication and we determine the moment when galaxies cross the observer past lightcone. For that, we make use of the galaxy merger trees provided by \LGalaxies{} which accurately follow in time the evolution of individual galaxies (and their corresponding \mBHS{} and \mbin{}) between DM snapshots with a time step resolution $\rm {\sim} 5\,{-}\,20\, Myr$ (see \citealt{IzquierdoVillalba2019LC} for further details). With this procedure we generate a full-sky lightcone with depth $z\,{\sim}\,4$ in which all the properties of galaxies, \mBHS{} and \mbin{} are stored. In Fig.~\ref{fig:LightCone} we present a slice of the full-sky lightcone used in this work, showing that galaxies  distributed in filaments and groups. To improve the readability we just plot the galaxies  with $\rm M_{stellar}\,{>}\,10^{10}\, M_{\odot}$ at $z\,{<}\,0.15$. In the figure we  highlight the galaxies hosting \mbin{} in an active phase, finding that active binaries tend to reside in over-dense regions of the Universe (as we will see in the following sections). We also point out three galaxies where we can find \mbin{} in different phases of their evolution: pairing phase (top panel), stellar hardening phase (top right panel) and gas hardening phase (lower right panel).

\section{Models for the growth for massive black holes and massive black hole binaries} \label{sec:GrowthModels}

In this work, we explore two different models for cold gas accretion onto the \mBHS{}. The first model (\fm{}) assumes that MBHs exhaust their gas reservoir in a \textit{fast} way. On contrary, the second one (\dm{}) relaxes that assumption allowing longer periods of gas accretion. These two different models allow us to explore how uncertainties in the gas consumption rate affect our predictions. In the following, we describe the two models:\\

(i) \textit{Fiducial model}: This model has been extensively used in \cite{IzquierdoVillalba2020} and \cite{IzquierdoVillalba2021}, providing a good match between predicted and observed quasar luminosity functions. As discussed in Section~\ref{sec:SAM_DESCRIPTIONS}, the main channel of MBH growth is the consumption of cold gas starting right after a galaxy merger or a disc instability episode. After these events, the cold gas available for accretion is assumed to settle in a reservoir around the black hole which is progressively exhausted according to a two phases model. The first phase corresponds to an Eddington-limited growth, which lasts until the MBH consumes a faction $\mathcal{F}$ of the available gas reservoir (see Eq.~\ref{eq:QuasarMode_Merger} and Eq.~\ref{eq:QuasarMode_DI}). $\mathcal{F}$ is a free parameter of the model and is set to 0.7 in order to match the faint end of the low-$z$ AGN LFs \citep[see e.g.][]{Marulli2008,Bonoli2009}. Once this phase ends, the BH enters in a self-regulated or quiescent growth regime characterized by progressively smaller accretion rates:
\begin{equation} \label{eq:feddQuiescent}
f_{\rm Edd} \,{=}\, \left[ 1 + (\mathit{t}/\mathit{t}_\mathit{Q})^{1/2}\right]^{-2/\beta},
\end{equation} 
\noindent where $f_{\rm Edd}$ is the accretion rate in Eddington units,  $t$ is the time referred to the starting time of the second phase and $t_Q \,{=}\, t_0\,\xi^{\beta}/(\beta \ln 10)$, being $t_0 \,{=}\, 1.26{\times}10^8 \, \rm yr $, $\beta \,{=}\, 0.4$ and $\xi \,{=}\, 0.3$. The choice of these values is based on the discussion presented in \cite{Hopkins2009} which showed that models of \textit{self-regulated} MBH growth require that $0.3\,{<}\,\beta\,{<}\,0.8$ and $0.2\,{<}\,\xi\,{<}\,0.4$. We highlight that the change of $\beta$ and $\xi$ values in the interval suggested by \cite{Hopkins2009} has a small effect on our results since the bulk of the MBH growth happens during the Eddington-limited phase.\\

\begin{figure}
\centering  
\includegraphics[width=1.0\columnwidth]{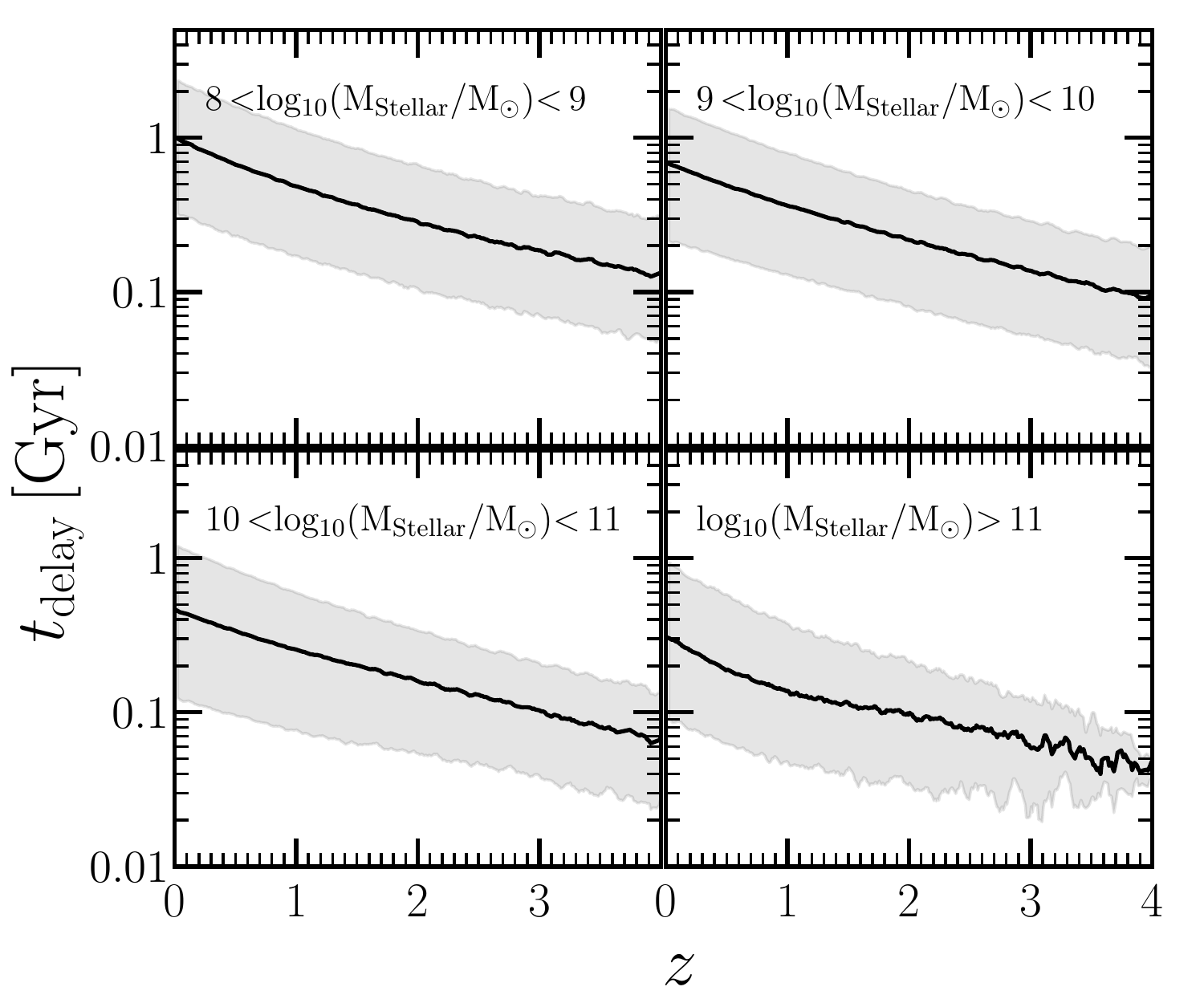}
\caption[]{Values of $t_{\rm delay}$ as a function of redshift for different stellar mass bins. The black lines represent the median value while grey areas display the $\rm 32^{th}-68^{th}$ percentile.}
\label{fig:tdelay}
\end{figure}

Regarding the growth of the binary, we follow \cite{IzquierdoVillalba2021} which assumes that the accretion rates of the two MBHs of the binary are correlated, as proposed by \cite{Duffell2020}: 
\begin{equation} \label{eq:Relation_accretion_hard_binary_blac_hole}
\dot{\rm M}_{\rm BH_1} =  \dot{\rm M}_{\rm BH_2} (0.1+0.9\mathit{q}),
\end{equation} 
where $q\,{=}\,\mathrm{M_{BH_2}/M_{BH_1}}\,{\leq}\,1$ is the binary mass ratio, $\dot{\rm M}_{\rm BH_1}$ and  $\dot{\rm M}_{\rm BH_2}$ are respectively the accretion rate of the primary and secondary MBH. As done in \cite{IzquierdoVillalba2021}, we fix the accretion rate of the secondary MBH to Eddington limit.\\

Finally, for MBHs in the dynamical friction phase, the growth is modeled in the same way as we do for nuclear black holes, i.e the accretion rate is determined by an initial Eddington limited phase followed by a self-regulated growth in which the black hole consumes the gas at low Eddington rates (see Eq.~\ref{eq:feddQuiescent}). The growth of MBHs in the dynamical friction phase ends when the black holes consume the total gas reservoir stored prior to the merger. This reservoir is set as the sum of all the gas that the MBH accumulated before the merger (i.e, as a consequence of disc instabilities or past mergers) and an extra amount computed at the time of the merger as Eq.~\ref{eq:QuasarMode_Merger}. This extra reservoir growth is motivated by the hydrodynamical simulations of merging galaxies with central MBHs by \cite{Capelo2015}. The authors showed that during the merging process the secondary galaxy suffers large perturbations during the pericenter passages around the central one. In these circumstances, the black hole of the secondary galaxy experiences accretion enhancements, mainly correlated with the galaxy mass ratio.\\

\noindent (ii) \textit{Delayed model}: In this model, we relax the fact that the growth of MBHs starts right after the galaxy merger. Instead, motivated by the theoretical work of \cite{HopkinsQuataert2010}, we assume that the gas needs several galaxy dynamical times before reaching the vicinity of the MBH and triggering the AGN activity. Therefore, we consider that the growth of the MBH is delayed  with respect to the galaxy merger by a time, $t_{\rm delay}$, given by:
\begin{equation}
t_{\rm delay} \,{=}\,n\,t_{\rm dyn}^{\rm gas}\,=\,n\,\frac{R_{\rm gas}}{\rm V_{max}} 
\end{equation}
where $R_{\rm gas}$ is the cold gas disc radius (see \citealt{Henriques2015} for further details), $\rm V_{max}$ is the maximum circular velocity of the DM halo and $n$ is a free parameter set to $3$. We have checked that the results presented here do not dramatically change when we double the value of $n$. Even though $n$ might vary with the galaxy merger ratio \citep[see e.g.][]{Hernquist1989,Naab2001}, we prefer to keep the model simple and we assume the same value of $n$ for all types of mergers. We highlight that we do not impose any $t_{\rm delay}$ after a disc instability given that the model assumes that the stars that build-up the bulge and the gas which fuels the \mBHS{} belong to the innermost part of the disc (see \citealt{Guo2011} and \citealt{Henriques2015} for further details). To give an idea about the typical gas delay included in the model, in Fig.~\ref{fig:tdelay} we present the evolution of $t_{\rm delay}$ as a function of redshift and stellar mass. As shown, regardless of galaxy mass, at $z\,{>}\,2$ the values rarely exceed $\rm 0.1\, Gyr$ while at $z\,{<}\,2$, $t_{\rm delay}$ increases up to $\rm 1\, Gyr$. As expected, when dividing the population in different stellar masses we can see that the smaller is the mass of the galaxy the larger is $t_{\rm delay}$. This is principally caused by the small maximum circular velocity of the halos in which these low-mass galaxies are hosted. \\

On top of the gas delay, in this model we relax the first phase of Eddington limited growth included in the \fm{}. Specifically, we assume that during that phase, $f_{\rm Edd}$ follows a log-normal distribution. Observational studies have suggested that at $z\,{<}\,3$, MBH follow a log-normal distribution centered at $0.3$ with a scatter of $0.25\,\rm dex$ \citep{Kollmeier2006}. However, when including these parameters in \LGalaxies, we did not find a good match with the observed luminosity functions at $z\,{>}\,2$. The model under-predicted the number of objects with $\rm L_{bol}\,{>}\,10^{46} \,erg/s$. To avoid this, we decide to treat the median and scatter values as free parameters. We found that the model was still able to match the bright end of the $z\,{>}\,2$ luminosity functions when a median and scatter of $0.8$ and $0.25\,\rm dex$\footnote{In case the extracted $f_{\rm Edd}$ has a value larger than $1$, we re-set it at $f_{\rm Edd}\,{=}\,1$} were used. Finally, this procedure applies also in the presence of an MBHB: we allow a delay between the merger and the gas accretion and we assume that the accretion rate of the secondary MBH follows the same log-normal distribution as single nuclear MBHs. For the case of MBHs in the dynamical friction phase, we do not apply any growth delay.\\

\begin{figure}
\centering
\includegraphics[width=1.\columnwidth]{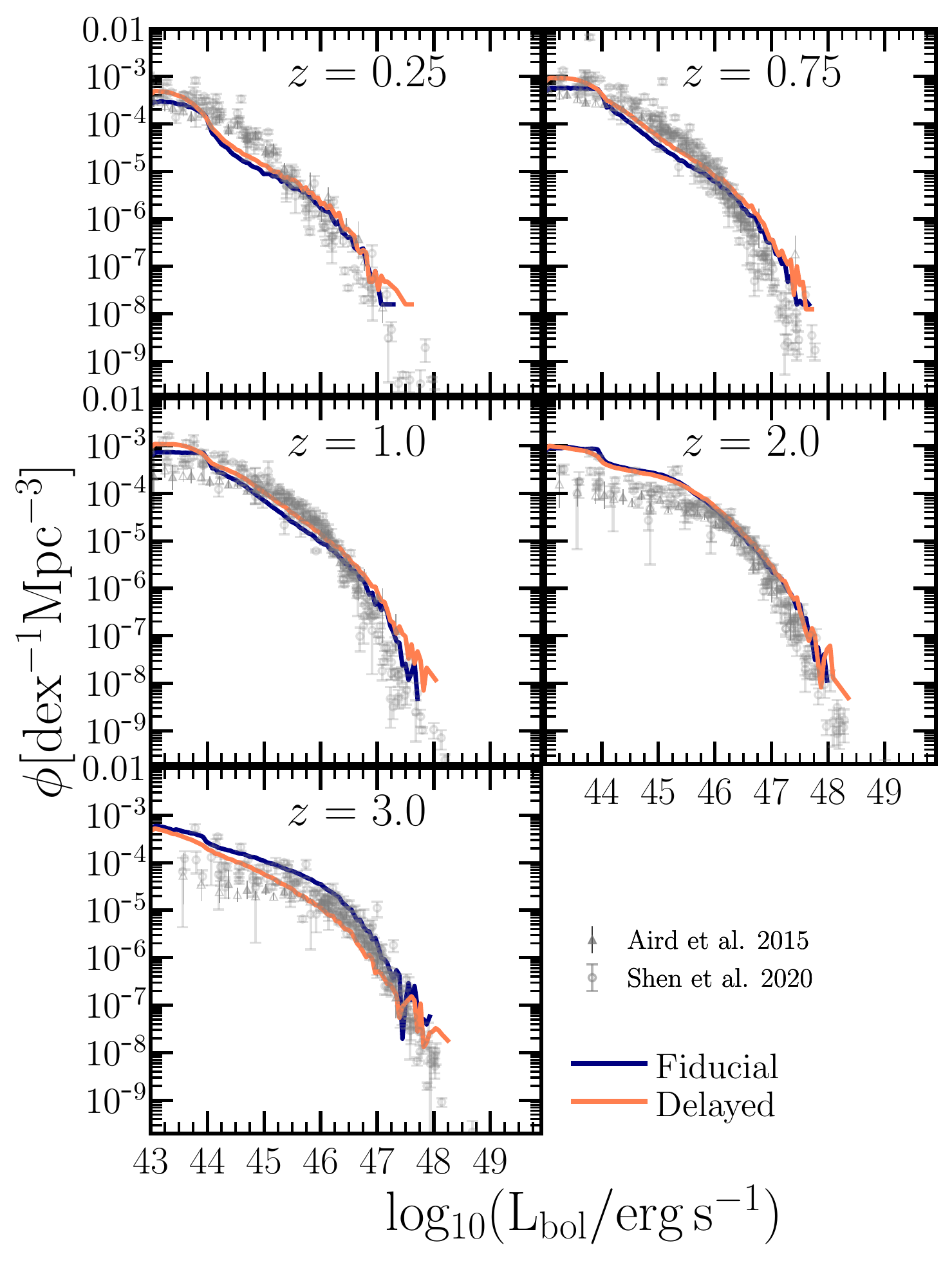}
\caption[]{Quasar bolometric luminosity functions ($\rm L_{bol}$) at $z\,{\approx}\,0.25, 0.75, 1.0, 2.0, 3.0$. Luminosity functions are compared with the data of \protect\cite{Shen2020} (circles) and \protect\cite{Aird2015} (triangles). In all the plots the blue and orange lines correspond to the predictions for the \textit{fiducial} and \textit{delayed} model, respectively.}
\label{fig:QSO_with_without_delay}
\end{figure}

To show that both \textit{fiducial} and \textit{delayed} models are able to reproduce the observed number density of active black holes, in Fig.~\ref{fig:QSO_with_without_delay} we present their predictions for the bolometric luminosity function at $z\,{<}\,3$. Both models are in agreement with the current constraints provided by \cite{Aird2015} and \cite{Shen2020}. At $z\,{>}\,2$, the main differences between the two models are seen in the bright end of the luminosity function ($\rm L_{bol}\,{>}\,10^{45}\, erg/s$). Specifically, the \dm{} displays a lower amplitude. This is caused by its smaller $f_{\rm Edd}$ values, which causes fainter objects at fix black hole mass and delays the assembly of \mBHS{} with respect to an Eddington limited growth. On the other hand, at lower redshifts both models converge at $\rm L_{bol}\,{>}\,10^{45}\, erg/s$ but they differ at lower luminosity, with the \dm{} resulting in slightly larger values. This is caused by the lower accretion rates allowed in the \dm{} which imply a slower consumption of the gas reservoir, thus resulting in an extended lifetime of the active phase.

\section{The environments of massive black hole binaries} \label{sec:HostsofMBHBs}
We now explore the properties of the hosts of
\mbin{} with $\rm M_{BH,1}\,{>}\,10^7\, M_{\odot}$. Specifically, we start by studying the stellar and halo content, we then move to the occupation fraction of \mbin{} and finally we examine the galaxy morphology. All these properties are explored on based of the binary mass ratio, $q\,{=}\, \rm {M_{BH,2}/M_{BH,1}}$. Here on, we denote as \textit{unequal}  binaries those having  mass ratio $q\,{<}\,0.1$, and \textit{equal} mass when $q\,{\geq}\,0.1$. We emphasize that by taking into account Eq.~\ref{eq:a0} (see also Fig.~\ref{fig:rho_a0_E_CB_PB}), the typical separation of the binaries studied here is $\rm {\lesssim}\,300\,pc$. Indeed, we have checked that our studied MBHB population displays a peak close to $1\,{-}\,10\,\rm pc$.

\subsection{Stellar and halo content of galaxies hosting massive black hole binaries}

\begin{figure}
\centering
\includegraphics[width=1.\columnwidth]{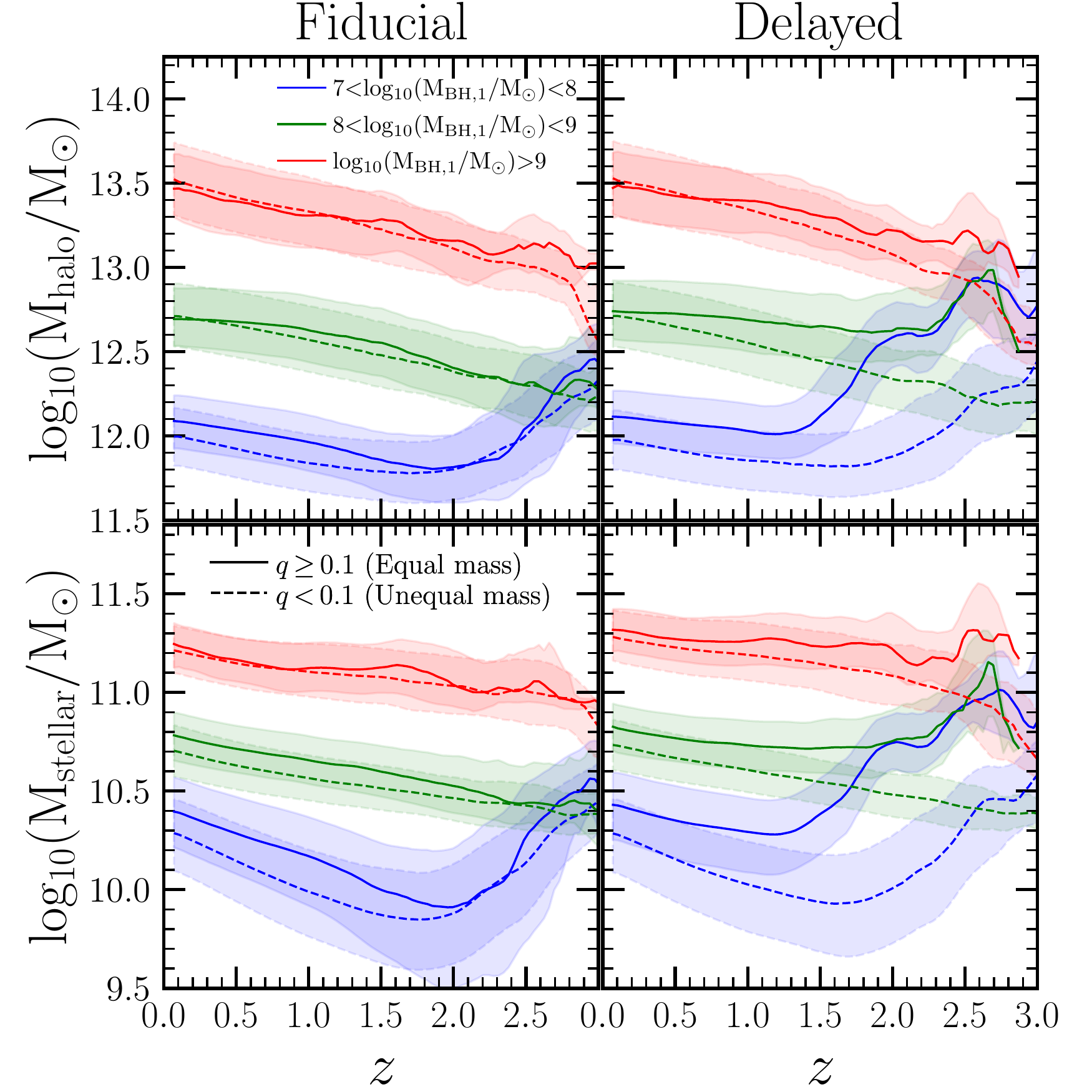}
\caption[]{Median halo ($\rm M_{halo}$) and stellar ($\rm M_{stellar}$) mass of the \mbin{} hosts. Shaded areas corresponds to the $32^{\rm th}\,{-}\,68^{\rm th}$ percentile. Each color represents a different mass bin: $\rm 7\,{<}\,log_{10}(M_{BH,1}/M_{\odot})\,{<}\,8$ (blue), $\rm 8\,{<}\,log_{10}(M_{BH,1}/M_{\odot})\,{<}\,9$ (green) and $\rm log_{10}(M_{BH,1}/M_{\odot})\,{>}\,9$ (red). In all the figures, solid and dotted lines correspond equal ($q\,{\geq}\,0.1$) and unequal ($q\,{<}\,0.1$) mass binaries, respectively. The left panel shows the results for the \fm{}, while the right panel does it for the \dm{}.}
\label{fig:MstellarMhalo_Binaires}
\end{figure}

\begin{figure}
\centering
\includegraphics[width=1.\columnwidth]{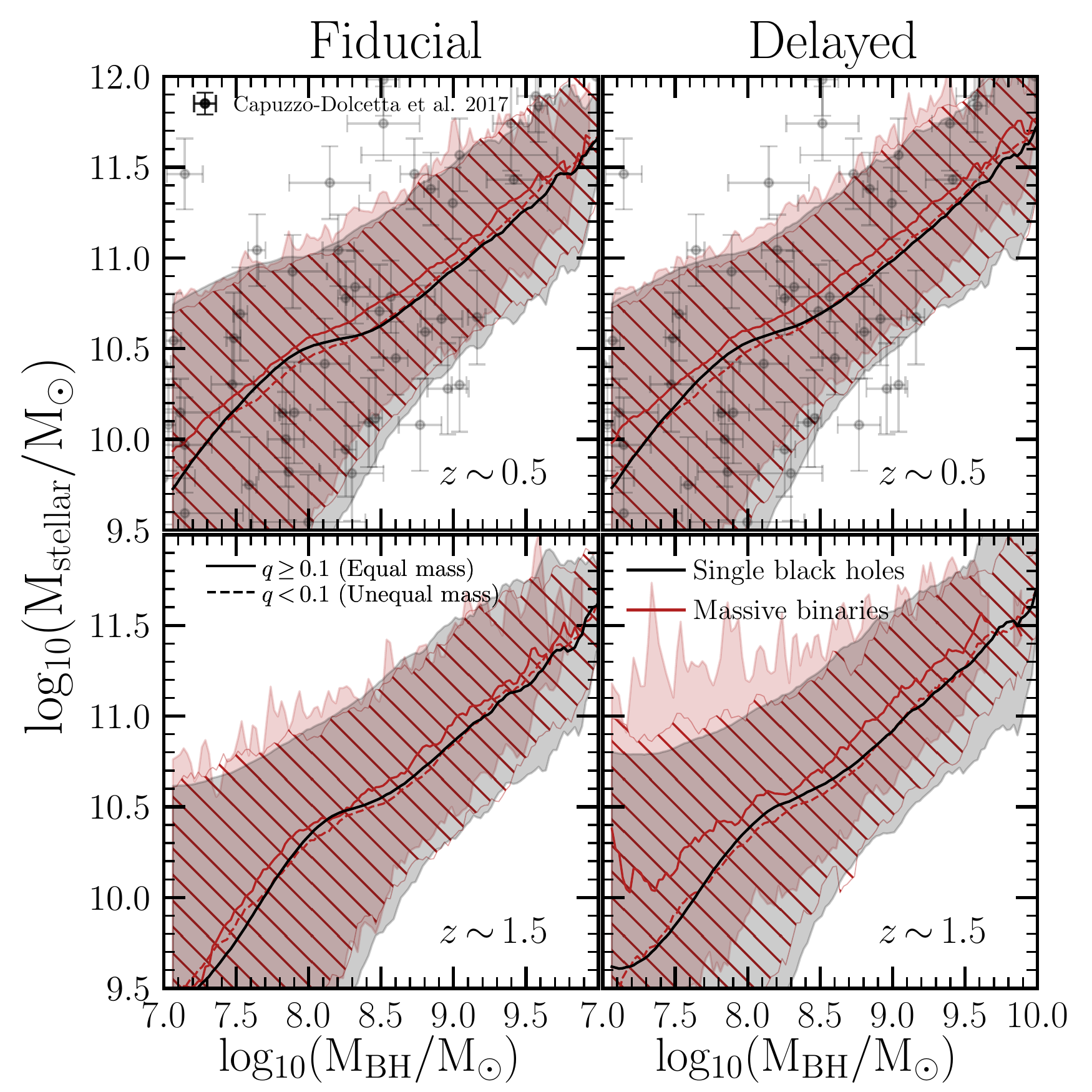}
\caption[]{Median stellar mass ($\rm M_{stellar}$) versus the black hole mass ($\rm M_{BH}$) for single (black colors) and binary black holes (red colors) at $z\,{\sim}\,1.5$ and $z\,{\sim}\,0.5$. For binaries, the value of $\rm M_{BH}$ corresponds to $\rm M_{BH,1}$. In all the panels, solid and dotted lines represent equal ($q\,{\geq}\,0.1$) and unequal ($q\,{<}\,0.1$) mass binaries, respectively. The black solid, red solid and red stripped areas correspond to the $10^{\rm th}\,{-}\,90^{\rm th}$ percentile of single MBHs, unequal and equal mass binaries, respectively. The left panel shows the results for the \fm{}, whereas the right panel does it for the \dm{}. For reference, we have added the observational points of \protect{\cite{Capuzzo2017}} at $z\,{=}\,0$. }
\label{fig:Mstellar_MBH}
\end{figure}

In Fig.~\ref{fig:MstellarMhalo_Binaires} we present the halo and the stellar mass of the galaxies hosting \mbin{}. As shown, regardless of the value of $q$ and growth model, binaries with $\rm M_{BH,1} \,{>}\,10^9\,M_{\odot}$ are placed in very massive galaxies and halos: $\rm 10^{11}\,{<}\,M_{stellar}\,{<}\,10^{11.5}\, M_{\odot}$ and $\rm 10^{13}\,{<}\,M_{halo}\,{<}\,10^{13.5}\, M_{\odot}$. In particular, the typical halo mass which hosts these \mbin{} evolves with redshift. While at $z\,{>}\,1.5$ it has a mass of $\rm {\sim}\,10^{13}\, M_{\odot}$, at $z\,{=}\,0$ the mass increases by $\rm 0.5\, dex$. On contrary, the stellar mass displays a milder evolution, reaching  a constant value of $\rm M_{stellar}\,{\sim}\,10^{11.25}\, M_{\odot}$ by $z\,{<}\,1.5$. Such different behavior is caused by the AGN feedback halting the stellar growth while leaving untouched the increase of the halo mass. For \mbin{} with  $\rm 10^8\,{<}\, M_{BH,1} \,{<}\,10^9\,\, M_{\odot}$, the typical mass of their hosts increases monotonically towards low-$z$, reaching by $z\,{\sim}\,0$ $\rm M_{halo}\, {\sim}\,10^{12.5}\, M_{\odot}$ and $\rm M_{stellar}\,{\sim}\,10^{10.75}\, M_{\odot}$. As happened for the most massive systems, no strong differences are seen when the population is divided in $q$ ratios. Finally, we note that binaries with $\rm 10^7\,{<}\, M_{BH,1} \,{<}\,10^8 \, M_{\odot}$  at $z\,{<}\,1$ are placed in Milky-Way type halos and galaxies: $\rm M_{halo}\,{\sim}\,10^{12}\, M_{\odot}$ and $\rm M_{stellar}\,{\sim}\,10^{10.5}\, M_{\odot}$. However, the evolution of these  quantities depends on the mass ratio $q$. In particular, unequal mass binaries are systematically placed in less massive halos and galaxies. Larger differences are seen in the stellar masses, where unequal mass binaries can be placed in systems up to $\rm 0.2\,dex$ less massive than their equal mass counterparts. Interestingly, this trend is more evident in the \dm{}, where the differences reaches up to $\rm {\sim}\,1\,dex$ at $z\,{>}\,1$. Such large deviation seen in the \dm{} is probably caused by the fact that MBHs forming equal mass binaries are typically assembled earlier than the ones placed in unequal mass systems. Thus, the earlier formation time of the former population leaves an imprint on the host in which it resides. Moreover, as shown in Fig.~\ref{fig:MstellarMhalo_Binaires},  galaxies hosting binaries of $\rm 10^7\,{<}\, M_{BH,1} \,{<}\,10^8 \, M_{\odot}$ are typically more massive at $z\,{\gtrsim}\,2.5$ than $z\,{=}\,0$. The trend shows that they are progressively less massive down to $z\,{\sim}\,2$, time at which this trend is reversed. This is caused by the fact that the process of the stellar and MBH assembly proceeds at different paces. At $z\,{>}\,3$ the hosts of $\rm {<}\,10^8\,M_{\odot}$ MBHs undergo a faster assembly than their nuclear MBHs. However, at $2\,{<}z\,{<}\,3$ the trend changes, and the growth of $\rm {<}\,10^8\, M_{\odot}$ MBHs takes place thanks to the fact that galaxies had enough time to develop massive discs able to induce important disc instabilities triggering the growth of MBHs but leaving untouched the galaxy assembly (disc instabilities in \LGalaxies{} only re-arrange the stellar mass between disc and bulge). This wiggling behavior is more evident in the \dm{}, especially for equal mass binaries. This is caused because the lagged growth of MBHs included in the \dm{} causes that the different paces at which galaxies and MBHs assembly become more extreme. Finally, we also see some differences between the stellar mass of the galaxy hosting equal and unequal mass binaries. This is caused by the type of galaxy mergers that brought the secondary MBH to the galaxy. While for equal mass binaries the secondary MBH is typically coming from mergers with large baryonic mass ratios, for unequal mass systems this is not the case. Since in \LGalaxies{} the burst of star formation after a galactic merger is proportional to mass ratio of the interacting galaxies, galaxy housing equal mass binaries experience a faster stellar assembly than galaxies with unequal mass systems. Such difference in the assembly is the main cause of the differences seen in Fig.~\ref{fig:MstellarMhalo_Binaires}.\\


To check if the \mbin{} follow a different scaling relation than single \mBHS{}, in Fig.~\ref{fig:Mstellar_MBH} we show at $z\,{\sim}\,1.5$ and $z\,{\sim}\,0.5$ the $\rm M_{BH}\,{-}\,M_{stellar}$ plane. For completeness we added the $z\,{=}\,0$ data of \cite{Capuzzo2017}. 
The results show that single and binary systems are placed in similar galaxies, regardless of redshift and growth model. When exploring the plane for unequal and equal mass systems, we can see that while the former does not show differences with respect to single \mBHS{}, the later is systematically placed in slightly more massive galaxies. This trend is particularly evident at $z\,{\sim}\,0.5$ in both models and at $z\,{\sim}\,1.5$ for the \dm{}. Nevertheless, these differences are small and the median relation of equal mass systems lies inside the scatter of single \mBHS{}.\\

\begin{figure}
\centering
\includegraphics[width=1.\columnwidth]{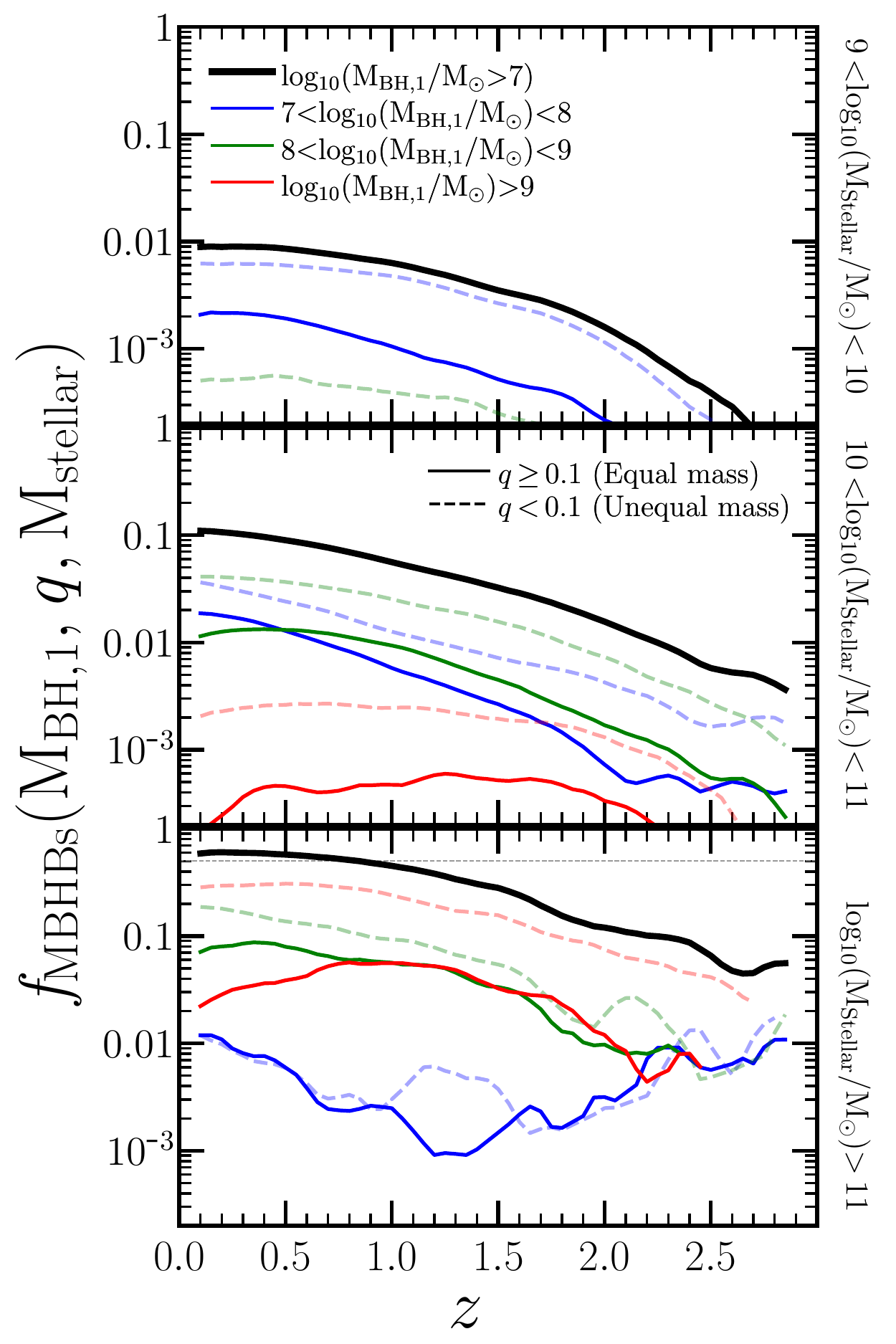}
\caption[]{Occupation fraction, $f_{\rm MBHBs}$, as a function of redshift and stellar mass for the \fm{}. Black color displays the population of MBHBs with $\rm M_{BH,1}\,{>}\,10^7\,M_{\odot}$ while colored ones refer to different mass bins: $\rm  7\,{<}\,log_{10}(M_{BH,1}/M_{\odot})\,{<}\,8$ (blue), $\rm 8\,{<}\,log_{10}(M_{BH,1}/M_{\odot})\,{<}\,9$ (green) and $\rm log_{10}(M_{BH,1}/M_{\odot})\,{>}\,9$ (red). In all the panels solid and dotted line corresponds to equal and unequal mass binaries, respectively.}
\label{fig:Ocupation_fraction_MBHB}
\end{figure}

As discussed above, Fig.~\ref{fig:MstellarMhalo_Binaires} and Fig.~\ref{fig:Mstellar_MBH} give us useful information about the type of galaxies which host \mbin{}, but it do not tell us how common such binaries are. For this, we explore the probability for a given galaxy to host a MBHB with $\rm M_{BH,1} \,{>}\,10^7 \, M_{\odot}$, hereafter occupation fraction ($f_{\rm MBHBs}$), which we show in in Fig.~\ref{fig:Ocupation_fraction_MBHB} . Here we define $f_{\rm MBHBs}$ as the ratio between galaxies hosting \mbin{} and the total number of galaxies. For the sake of simplicity, we only show the results for the \fm{} given that we have checked that both growth models explored here display a similar behaviour. Looking at the upper panels of the figure, galaxies with $10^9{<}\,\rm M_{stellar}\,{<}\,10^{10}\, M_{\odot}$ rarely host a \mbin{} with $\rm M_{BH,1} \,{>}\,10^7 \, M_{\odot}$, regardless of redshift. Although the probability increases towards low redshifts, it never exceeds 1\%. When dividing the population in different mass bins and $q$ values, most of the binaries hosted in these galaxies are unequal mass ($q\,{<}\,0.1$) with masses $\rm 10^7\,{<}\,M_{BH,1}\,{<}\,10^8\, M_{\odot}$. We highlight that this small occupation fraction is just a consequence of the adopted mass threshold ($\rm{>}\,10^7\,M_{\odot}$) and not because  \mbin{} are rarer in these galaxies. Indeed, by reducing the mass of the primary black hole down to $\rm 10^6\, M_{\odot}$, the occupation fraction increases ${\sim}\,1.5\, \rm dex$. For galaxies with $\rm 10^{10}\,{<}\,\rm M_{stellar}\,{<}\,10^{11}\, M_{\odot}$ we see substantial changes, as the occupation fraction increases by a factor $10$ in both the \textit{fiducial} and the \textit{delayed} models. For instance, $f_{\rm MBHBs}$ reaches up to 10\% by $z\,{\sim}\,0$, although it remains below 1\% at $z\,{>}\,2$. When dividing the population in mass and $q$, we can appreciate that the evolution of $f_{\rm MBHBs}$ is dominated  by unequal mass binaries with $\rm 10^7 {<}\, M_{BH,1} \,{<}\,10^9 \, M_{\odot}$. Equal mass binaries are rarer, being present at most in $\approx 2\%$ of the galaxies at $z<1$. Finally, the lower panel of Fig.~\ref{fig:Ocupation_fraction_MBHB} shows $f_{\rm MBHBs}$ for the most massive galaxies in the lightcone, ($\rm M_{stellar}\,{>}\,10^{11}\, M_{\odot}$). For these systems, by $z\,{<}\,1$ the model predicts a whopping $f_{\rm MBHBs}\,{\sim}\,50\%$. The typical binaries hosted in these galaxies have $q<0.1$ with $\rm M_{BH,1} \,{>}\,10^9 \, M_{\odot}$. However, up to 10\% of these galaxies can host equal mass \mbin{} with primary mass of $\rm M_{BH,1} \,{>}\,10^8 \, M_{\odot}$. These results highlight that, perhaps contrary to conventional wisdom, bound MBHBs are very common in massive galaxies at low redshift. In practice, every other massive galaxy at $z<1$ should host a binary with separation $<300\,$pc, and one out of ten should host a relatively equal mass binary. Therefore, massive low redshift galaxies should be primary targets for searching MBHBs, although as we will discuss later on that only a small fraction of them is likely to display significant electromagnetic signatures.\\

Although for low-$z$ massive galaxies the occupation fraction of MBHBs is relatively large, the model shows that many galaxies do not host a binary. This is probably caused by a combination of different reasons. On one hand, after a merger, many satellite galaxies deposit small MBHs ($\rm {<}\,10^6\, M_{\odot}$) whose dynamical friction phase can easily last longer than the Hubble time, leaving these MBH wandering in the galaxy. As a consequence, not all galaxy encounters are efficient in depositing MBHs that can potentially lead to the formation of parsec scale MBHBs. On the other hand, the coalescence of MBHs is another channel through which the occupation fraction can decrease. This is especially important for the mass range we are exploring. For instance, systems with $q\,{>}\,0.1$ and $\rm M_{BH,1}\,{>}\,10^7\, M_{\odot}$ have a hardening phase that lasts ${<}\,1\,\rm Gyr$. Finally, the quiet merger history of some galaxies hampers the formation of MBHBs. This is the case of $z\,{=}\,0$ galaxies hosting a pseudobulge structure. These systems represent at $z\,{=}\,0$ more than 60\% of the galaxies with $\rm M_{stellar}\,{>}10^{10}\, M_{\odot}$ but less than 40\% (none) of them experience a minor (major) merger (see Figure 16 of \citealt{IzquierdoVillalba2019}).

\begin{figure}
\centering
\includegraphics[width=1.\columnwidth]{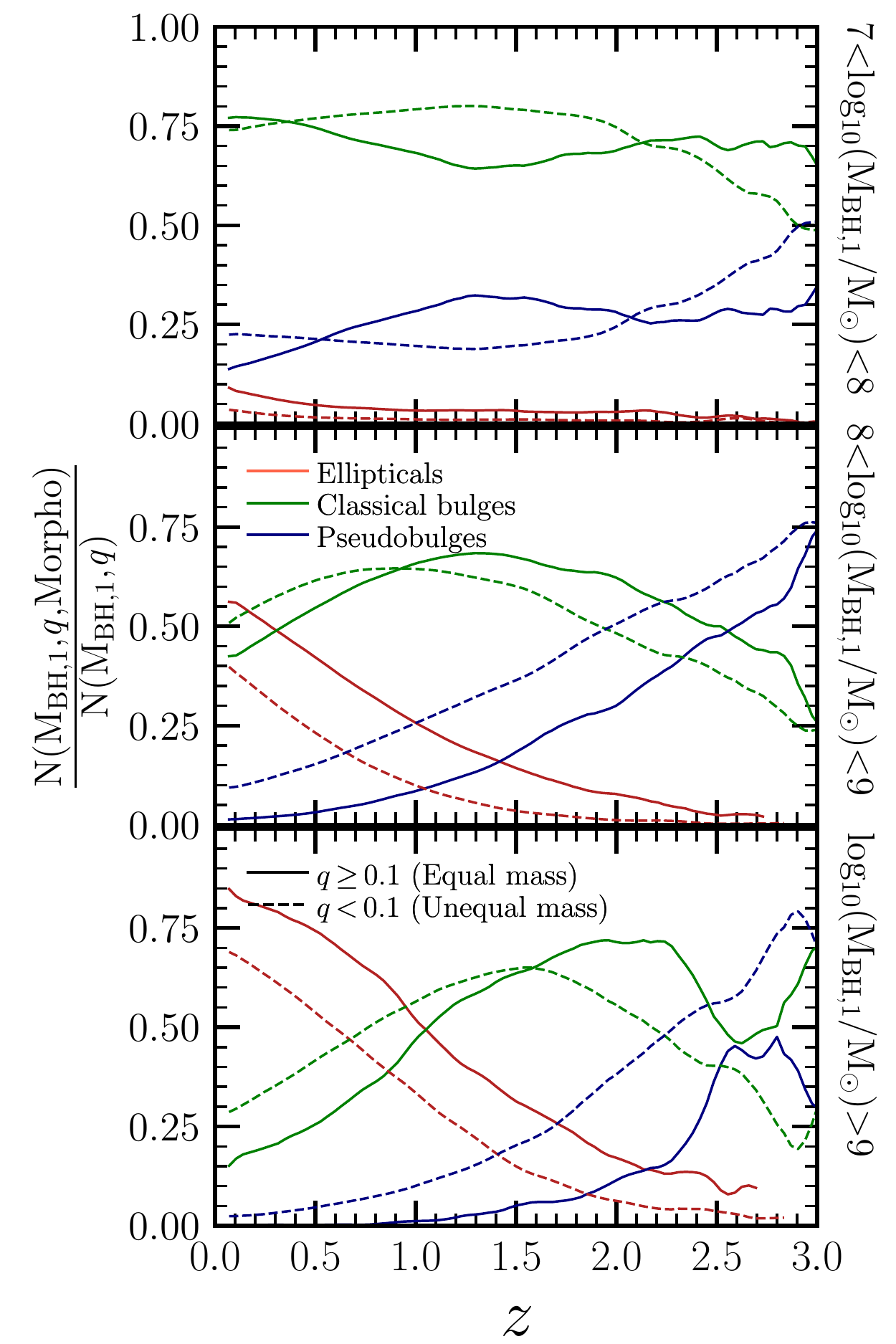}
\caption[]{Redshift evolution of the morphology of galaxies hosting \mbin{}. Red, green and blue color correspond to elliptical-, classical- and pseudo- bulge morphology Upper, middle and lower panel represent the results for  $\rm   7\,{<}\,log_{10}(M_{BH,1}/M_{\odot})\,{<}\,8$, $\rm 8\,{<}\,log_{10}(M_{BH,1}/M_{\odot})\,{<}\,9$ and $\rm log_{10}(M_{BH,1}/M_{\odot})\,{>}\,9$, respectively. In all the panels solid and dotted lines correspond to equal and unequal mass binaries, respectively. We only show the results for the \fm{} given that the \textit{delayed} one displays similar trends.}
\label{fig:Morphology_Hosts_of_Binaries}
\end{figure}

\subsection{The morphology of galaxies hosting massive black hole binaries}

Once studied the mass of the MBHB hosts, we explore the redshift evolution of their morphology. As we did before, to avoid crowded plots we only present the results for the \fm{} since both growth models display similar trends. From hereafter, we define as elliptical (classical bulge) any galaxy with $\rm B/T{>}\,0.7$ ($\rm 0.01\,{<}\,B/T\,{<}\,0.7$), where $\rm B/T$ is the bulge-to-total stellar mass ratio. On the other hand, a galaxy is tagged as pseudobulge when 2/3 of the mass inside the bulge was assembled trough disc instabilities \citep[see][for more details]{IzquierdoVillalba2019}.\\

In the upper panel of Fig.~\ref{fig:Morphology_Hosts_of_Binaries} we present the morphology of the galaxies hosting $\rm 10^7\,{<}\, M_{BH,1} \,{<}\,10^8 \, M_{\odot}$. Regardless of redshift and growth model, these binaries rarely inhabit elliptical structures, with less than 10\% of the cases at any redshift. The large majority of these binaries (${>}\,70\%$) are hosted in spiral galaxies whose nuclear component is characterized by a classical bulge. The other 20\% of the cases corresponds to spiral galaxies with a pseudobulge. Interestingly, this trend is maintained when the population is divided between equal and unequal mass binaries. \mbin{} with $\rm 10^8\,{<}\, M_{BH,1} \,{<}\,10^9 \, M_{\odot}$ display similar trends, being spiral galaxies with a classical bulge the preferred hosts at any redshifts and $q$ value. Nevertheless two differences can be appreciated: elliptical hosts become fairly common at $z\,{<}\,1$, contributing to 25\%-50\% of the cases, and pseudobulges host ${>}\,50\%$ of binaries at $z\,{>}\,2$. This is because in \LGalaxies{} disc instabilities are the main mechanism for bulge growth at $z\,{>}\,2$ \citep{IzquierdoVillalba2020}. Therefore, disc instabilities taking place after a minor merger responsible of the MBHB formation can trigger the development of a pseudobulge, erasing the pre-existent classical bulge structure. A similar process can occur in elliptical galaxies thanks to the fast regeneration of stellar disc at high-$z$. Finally, \mbin{} with $\rm M_{BH,1} \,{>}\,10^9 \, M_{\odot}$ are mainly placed in elliptical and classical bulge but with a different proportion depending on the redshift considered. While at $z\,{<}\,1$ the preferred morphology is the elliptical structure ($\,{>}\,50\%$ of the cases), at $z\,{>}\,1$ classical bulges take the main role.\\

\begin{figure}
\centering
\includegraphics[width=1.\columnwidth]{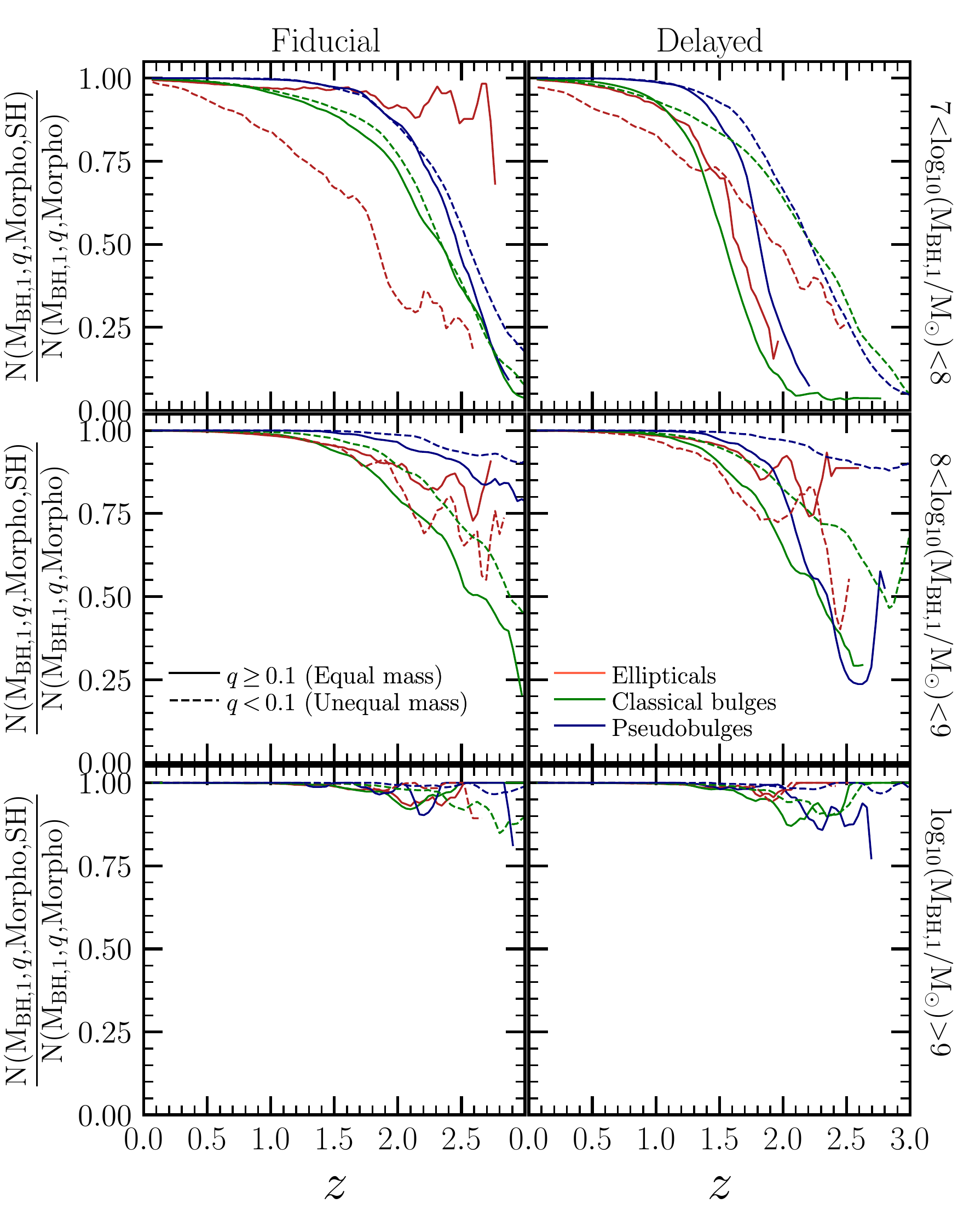}
\caption[]{Redshift evolution of the fraction of \mbin{} shirking due to stellar hardening. Red, green and blue lines correspond to elliptical-, classical- and pseudo- bulge morphology. Upper, middle and lower panel represent the results for  $\rm   7\,{<}\,log_{10}(M_{BH,1}/M_{\odot})\,{<}\,8$, $\rm 8\,{<}\,log_{10}(M_{BH,1}/M_{\odot})\,{<}\,9$ and $\rm log_{10}(M_{BH,1}/M_{\odot})\,{>}\,9$, respectively. The left panels display the results for the \fm{} while the right panels do it for the \dm{}. In all the panels solid and dotted lines correspond to equal and unequal mass binaries, respectively.}
\label{fig:Morphology_Stellar_Hard_Hosts_of_Binaries}
\end{figure}

As discussed in Section~\ref{sec:MBH_and_MBHBHs_Model}, binaries hosted in the galactic nucleus are able to reduce their separation through the interaction with a massive circumbinary gaseous disc (gas hardening) or by three-body interactions with single stars (stellar hardening). Given that the model links the formation of each bulge type with different formation scenarios (major/minor mergers or secular disc instabilities), it is interesting to study if the bulge morphology traces different environments around the \mbin{}, and thus different processes in which binaries reduce their separation. To explore which shrinking mechanism is the predominant in each galaxy morphology, in Fig.~\ref{fig:Morphology_Stellar_Hard_Hosts_of_Binaries} we computed the redshift evolution of the fraction of binaries shrinking though stellar hardening. Despite distinct assumptions on the timing and duration of gas accretion, no strong differences are seen between the \textit{fiducial} and \textit{delayed model}. However, the predictions depend on the mass of the primary black hole. Regardless of redshift and morphology, ${\sim}\,100\%$ of the binaries with $\rm M_{BH,1}\,{>}\,10^9\, M_{\odot}$ reduce their separation through encounters with stars. This is the consequence of the fact that these \mbin{} are hosted in the most massive galaxies of the simulation. Such systems are characterized by small gas fractions since they consumed most of the gas during the assembly of the stellar component \citep[consistent with][]{vanDokkum2005,Lin2008,Lin2010} and further gas cooling events were efficiently suppressed by the activity of the central massive black hole. In this way, most of the mergers that lead to the formation of \mbin{} with $\rm M_{BH,1}\,{>}\,10^9\, M_{\odot}$ are gas poor, preventing the formation of massive circumbinary discs around the MBHB. This changes when we decrease $\rm M_{BH,1}$. For $\rm 10^8\,{<}\,M_{BH,1}\,{<}\,10^9\,M_{\odot}$  we can see that at $z\,{>}\,2$ around $\rm 50\,{-}\,70\%$ of the binaries hosted in elliptical galaxies and classical bulges (with slightly smaller values in the \dm{}) reduce $a_{\rm BH}$ through stellar hardening. Interestingly, in pseudobulge structures, this percentage increases to $90\%$, a consequence of the fact that their binary systems were formed ${\sim}\,0.5\, \rm Gyr$ before the ones in elliptical and classical bulges and thus, had more time for consuming the gas reservoir and enter the stellar-driven hardening. At $z\,{<}\,1$ the large majority of the population shrinks via stellar encounters, regardless of bulge morphology. The increasing weight of stellar hardening towards low redshifts is caused by two effects. The first one is that at $z\,{<}\,1$ the hosts of these \mbin{} are massive enough to experience a similar gas reduction than the ones of $\rm M_{BH,1}\,{>}\,10^9\, M_{\odot}$ \mbin{}. The second effect is caused by the fact that circumbinary discs progressively decrease their mass (and eventually disappear) due to binary mass accretion. These trends are confirmed for $\rm 10^7\,{<}\,M_{BH,1}\,{<}\,10^8\, M_{\odot}$, where differences between high and low redshift are even larger. At $z\,{>}\,2$, less than half of the binaries are primarily driven by stellar hardening. The exact percentage varies depending on the galaxy morphology. For instance pseudobulge structures display values ${\sim}\,50\%$ whereas classical bulge have values ${\sim}\,20\%$. When comparing \textit{fiducial} and \textit{delayed} model, we can see that the latter predicts smaller fractions at high-$z$. This is primarily because in low mass galaxies the lagged gas consumption imposed by the \dm{} makes it more likely that a significant amount of cold gas is still in place when the MBHB forms. This gas forms a massive circumbinary disc that dominates the dynamical evolution of the binary over stellar scattering. Besides this, $\rm 10^7\, M_{\odot}{<}\,M_{BH,1}\,{<}\,10^8\, M_{\odot}$ equal mass binaries hosted in high-$z$ elliptical galaxies display strong differences between the \textit{fiducial} and \textit{delayed} model. We have checked that this is caused by the stellar masses of these structures, being up to $1.5\,\rm dex$ more massive in the \dm{} than in the \textit{fiducial} one (${\sim}\,10^{10}\, \rm M_{\odot}$ versus ${\sim}\,10^{8.5}\, \rm M_{\odot}$, respectively). Therefore, at a fixed black hole mass, the larger the stellar content of a high-$z$ galaxy, the larger its cold gas mass, and the larger the amount of matter that can be supplied onto the MBH after a merger or disc instability (see Eq.~\ref{eq:QuasarMode_Merger} and Eq.~\ref{eq:QuasarMode_DI}). On the other hand, we do not find these large differences in $\rm 10^7\, M_{\odot}{<}\,M_{BH,1}\,{<}\,10^8\, M_{\odot}$ equal mass binaries hosted in high-$z$ pseudobulges. This is because these bulge structures are placed in a similar type of galaxies in both versions of the model.

\section{Massive black hole binaries in the sky: Redshift distribution and electromagnetic signatures}  \label{sec:ElectromagneticCounterpart}
In this section, we explore redshift evolution of the number of massive black hole binaries and the fraction of those showing an electromagnetic signature. Besides, we study the occurrence of an active phases in \mbin{} and examine if these active systems can be distinguished from quasars/AGNs triggered by single \mBHS{}.

\subsection{The redshift distribution of massive black hole binaries}

\begin{figure}
\centering
\includegraphics[width=1.\columnwidth]{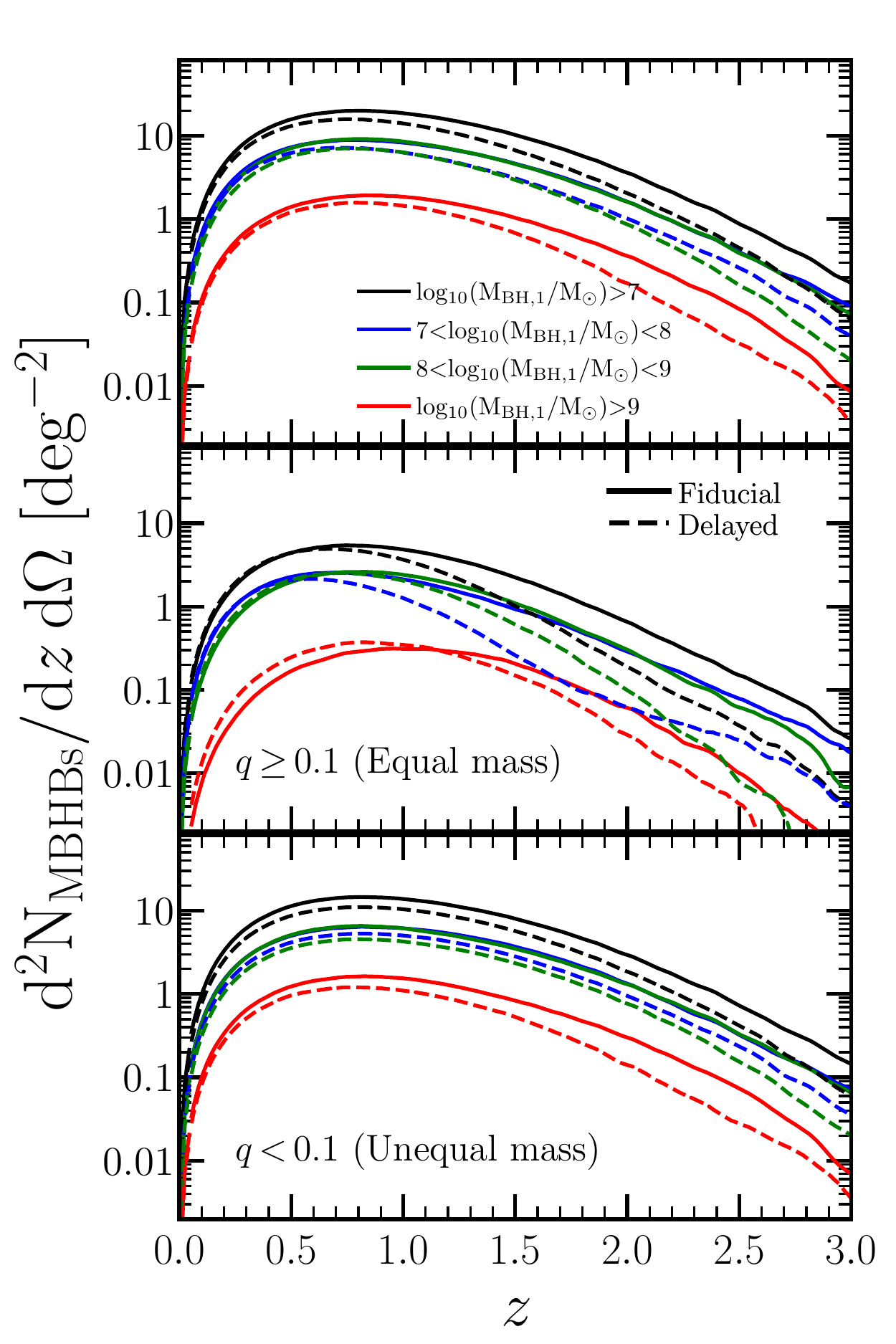}
\caption[]{Number of massive black hole binaries per square degree ($\rm deg^2$) for the \textit{fiducial} and \textit{delayed} model (solid and dashed lines, respectively). Black color displays the population of MBHBs with $\rm M_{BH,1}\,{>}\,10^7\,M_{\odot}$ while colored ones refer to different mass bins: $\rm  7\,{<}\,log_{10}(M_{BH,1}/M_{\odot})\,{<}\,8$ (blue), $\rm 8\,{<}\,log_{10}(M_{BH,1}/M_{\odot})\,{<}\,9$ (green) and $\rm log_{10}(M_{BH,1}/M_{\odot})\,{>}\,9$ (red). The two lower panels represent the same but dividing the population between equal and unequal mass binaries. The bin size of the redshift distribution corresponds to $0.01$.}
\label{fig:Number_density_MBHB}
\end{figure}


\begin{figure}
\centering
\includegraphics[width=1.\columnwidth]{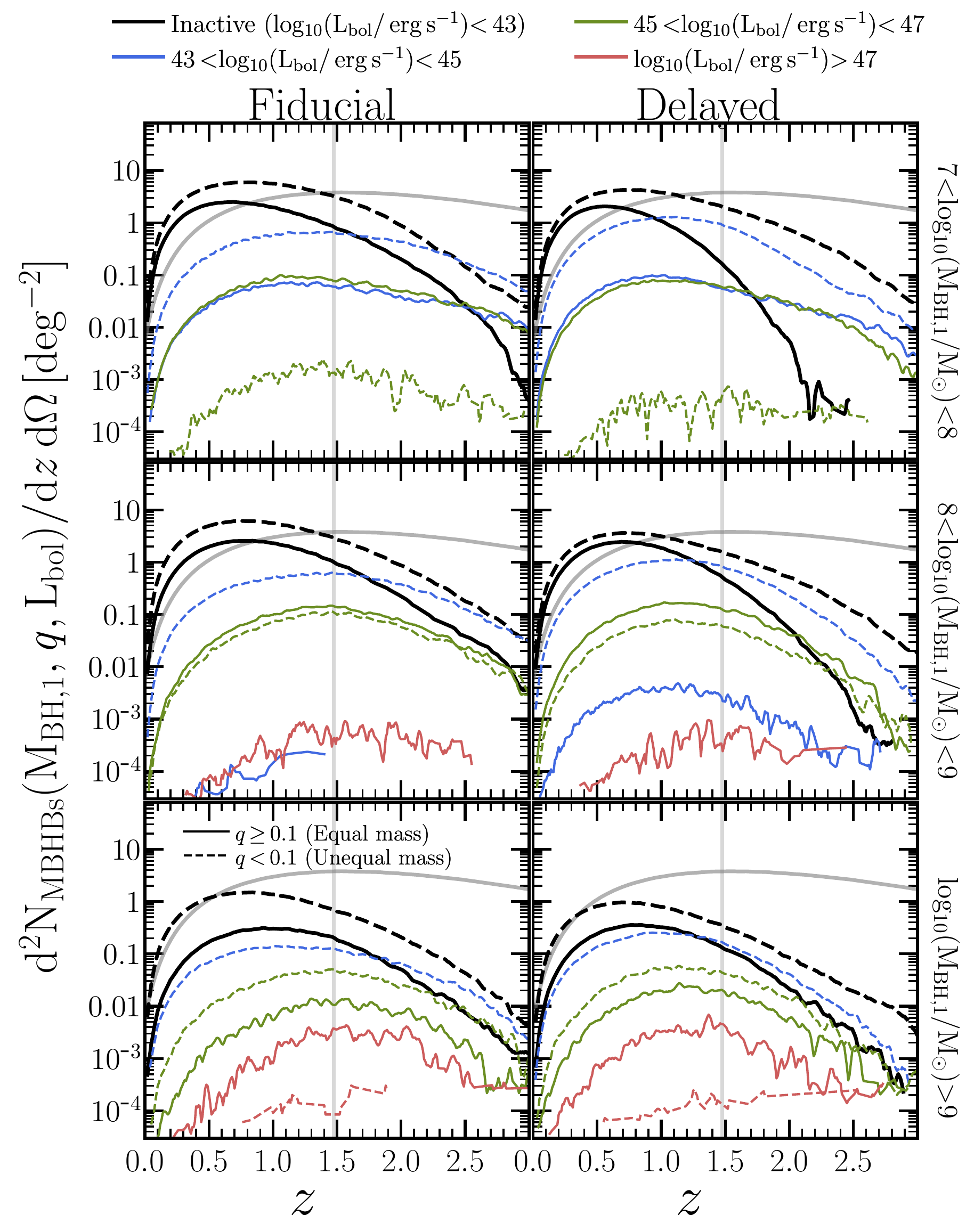}
\caption[]{Redshift distribution of the number of binaries per square degree radiating at different bolometric luminosity bins. Upper right, lower left and lower right panels represent the results for $\rm 7\,{<}\,log_{10}(M_{BH,1}/M_{\odot})\,{<}\,8$, $\rm 8\,{<}\,log_{10}(M_{BH,1}/M_{\odot})\,{<}\,9$ and $\rm log_{10}(M_{BH,1}/M_{\odot})\,{>}\,9$ binaries, respectively. In all panels, solid and dotted lines correspond to equal and unequal mass binaries, respectively. Left and right figures show the results for the \textit{fiducial} and \textit{delayed model}, respectively. To guide the reader in all the panels  the grey line shows  the redshift evolution of the number of galaxy mergers per square degree observed inside the lightcone, The vertical grey line highlights the peak of this latter distribution. The bin size of the redshift distribution corresponds to $0.01$.}
\label{fig:Number_density_Active_MBHB}
\end{figure}

From an observational point of view, one of the first questions that arises when studying \mbin{} is at which redshift the probability of finding these systems is the largest. To shed light on this question, in Fig.~\ref{fig:Number_density_MBHB} we present the predicted number of dynamically bound MBHBs per $\rm deg^2$ at different redshifts, $N_{\rm MBHBs}$. Independently of the growth model, \LGalaxies shows an increasing trend until $z\,{\sim}\,0.8$, where up to $\rm {\sim}\,20 \, objects$ can be found in $1\, \rm deg^2$. After $z\,{\sim}\,0.8$, the expected number drops very quickly, 
 vanishing at $z\,{\sim}\,0$. When dividing the population in different mass bins, we can see that binaries with $\rm 10^7\,{<}\,M_{BH,1}\,{<}\,10^8\, M_{\odot}$ and $\rm 10^8\,{<}\,M_{BH,1}\,{<}\,10^9\, M_{\odot}$ display very similar $N_{\rm MBHBs}$ values. However, systems with $\rm M_{BH,1}\,{>}\,10^9\, M_{\odot}$ are rarer and the largest probability of finding them is at $z\,{\sim}\,0.8$ with $\rm {\sim}\,1 \, object$ per $\rm deg^2$. In the same plot we have divided the population between equal ($q\,{\geq}\,0.1$) and unequal mass ($q\,{<}\,0.1$) binaries.  Irrespective of the mass and model, unequal mass binaries dominate the population, being typically $2\,{-}\,3$ times more common. Interestingly, while the predictions for unequal mass systems are not strongly affected by the growth model, in the case of equal mass binaries we find some differences. Specifically, at high redshift ($z\,{>}\,1$) they are less frequent in the \dm{}, with the largest differences affecting the $\rm 10^7\,{<}\,M_{BH,1}\,{<}\,10^8\, M_{\odot}$ and $\rm 10^8\,{<}\,M_{BH,1}\,{<}\,10^9\, M_{\odot}$ populations. For instance, the \dm{} predicts $5\,{-}\,7$ times less equal mass binaries at $z\,{>}\,1$. 
This difference is a combination of several factors. On one hand, as a result of the lagged growth included in the \dm{}, MBHs deposited after a galaxy-galaxy merger are typically less massive in this model than in the \textit{fiducial} one (we stress that the growth of massive black holes is delayed for the whole population of central MBH/MBHBs). Therefore, as a result of their lower mass, MBHs undergoing the dynamical friction phase in the \dm{} take more time to reach the galaxy nucleus (and often stall before getting there). Over this time, gas is fed to the nuclear MBHs triggering their mass growth. Thus, when the secondary MBH exits the dynamical friction phase, the mass difference between the nuclear and the incoming MBH is larger in the \dm{} than in the \textit{fiducial}. On the other hand, MBHs that start the dynamical friction phase are allowed to accrete the gas reservoir they had in the previous galaxy (i.e when they were centrals). It is common that these reservoirs are smaller in the \dm{} since at the moment of the merger most of the gas that would end up around the MBH is still infalling towards the center of the galaxy. As a result, MBHs in the \dm{} will grow less during the pairing phase. The combination of these factors cause that, at the binary formation time, the mass ratio distribution of the \dm{} is more skewed towards low $q$-values than the one predicted by the \fm{}. Finally, growth events taking place after binary formation lead to a reduction in the mass difference (see Section\ref{sec:MBH_and_MBHBHs_Model}). However, in the \dm{} these events are lagged causing that the binary systems require more time to increase their mass ratio.\\


\subsection{The electromagnetic signatures of massive black hole binaries}

Even though the previous section provides useful information about the redshift distribution of \mbin{}, it is fundamental to explore how it changes when we only consider systems displaying electromagnetic signatures. In this section we refer to $\rm L_{bol}$ as the total luminosity of the system computed as the sum of the primary and secondary black hole luminosity. This choice is motivated by the fact that the projected angular separation of our \mbin{} (${<}\,0.01\, \rm arcsec$) hampers the detection of dual AGNs by current electromagnetic observatories. In Fig.~\ref{fig:Number_density_Active_MBHB} we present the number of active binaries per square degree radiating at different bolometric luminosity. Regardless of the mass and $q$ parameter the \fm{} shows a maximum at $z\,{\sim}\,1.2-1.4$ (depending on the exact mass and luminosity). By exploring the model predictions at different mass bins, we can see that most of the active population with  $\rm 10^7\,{<}\,M_{BH,1}\,{<}10^8\, M_{\odot}$ displays $\rm 10^{43}\,{<}\,L_{bol}\,{<}\,10^{45}\, erg/s$ with ${\sim}\,0.5$ objects per $\rm deg^2$ at $z\,{\sim}\,1.5$. This number decreases quickly towards low-$z$, reaching by $z\,{\sim}\,0.1$ less than $10^{-3}$ accreting binary per $\rm deg^2$. Interestingly, the \mbin{} triggering this faint population are mostly unequal mass binaries. The largest contribution for equal mass systems is found at $\rm 10^{45} {<} L_{bol}\,{<}\,10^{47}\, erg/s$ but it is a much rarer population with ${\lesssim}\,0.08$ systems per $\rm deg^2$ at $z\,{\sim}\,1.5$. Similar trends are found in the $\rm 10^8\,{<}\,M_{BH,1}\,{<}10^9\, M_{\odot}$ population where most active binaries correspond to unequal mass systems radiating at $\rm L_{bol}\,{<}\,10^{45}\, erg/s$. Brighter objects are less common (by ${\sim}\,1\,$dex) with the large majority being powered by equal mass systems. Finally, active binaries with $\rm M_{BH,1}\,{>}10^9\, M_{\odot}$ are ${\sim}\,1\, \rm dex$ less frequent that those in the other mass bins. Despite this, the model shows that such kind of binaries are able to power very bright quasars ($\rm L_{bol}\,{>}\,10^{47}\, erg/s$) at $z\,{\sim}\,2$ but their occurrence drops rapidly after $z\,{\sim}\,1$.\\

\begin{figure}
\centering
\includegraphics[width=1.\columnwidth]{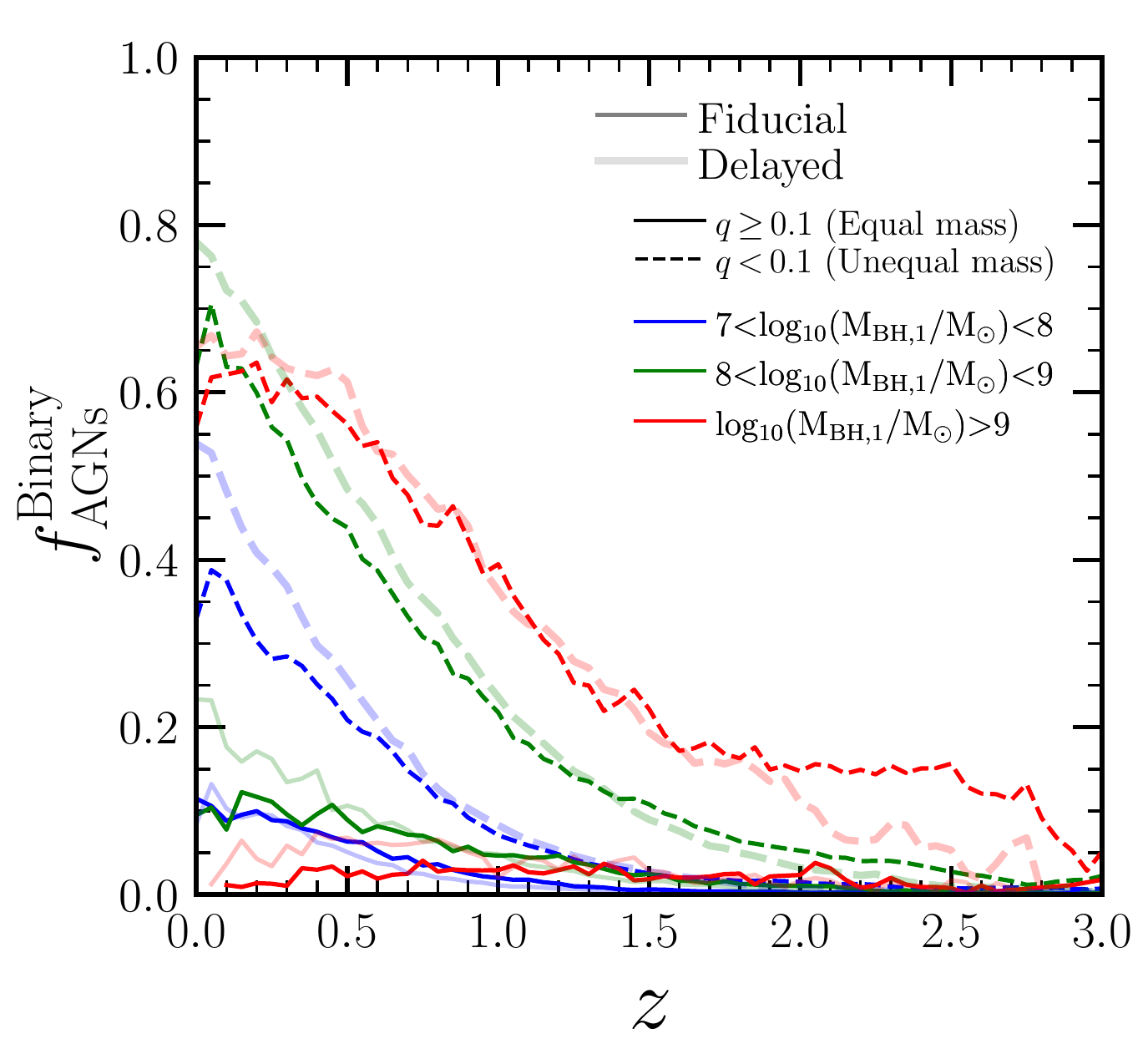}
\caption[]{Redshift evolution of the probability that a given observed quasar/AGN is powered by a binary system ($f_{\rm AGNs}^{\rm Binary}$). Each curve represent $f_{\rm AGNs}^{\rm Binary}$ for different black hole populations: $\rm 7\,{<}\,log_{10}(M_{BH,1}/M_{\odot})\,{<}\,8$ (blue), $\rm 8\,{<}\,log_{10}(M_{BH,1}/M_{\odot})\,{<}\,9$ (green) and $\rm log_{10}(M_{BH,1}/M_{\odot})\,{>}\,9$ (red). In all panels, solid and dotted lines correspond to equal and unequal mass binaries, respectively. Solid and transparent lines represent the results for the \textit{fiducial} and \textit{delayed model}, respectively.}
\label{fig:FAGNs_Binaries}
\end{figure}

The predictions for the \dm{} are presented in the right panels of Fig.~\ref{fig:Number_density_Active_MBHB}. As shown, the peak of the \mbin{} activity does not coincide with the one of seen in the fiducial model. The former happens $\rm {\sim}\,1\, Gyr$ after the latter. Besides this, the shape of the distribution has strong differences with respect to the one shown for the \fm{}. In particular, the activity of \mbin{} is heavily suppressed at high redshifts and enhanced at lower ones. By exploring the predictions as a function of mass, the behaviour for $\rm 10^7\,{<}\,M_{BH,1}\,{<}10^8\, M_{\odot}$ is similar to the one shown by the \textit{fiducial} one: faint and unequal binaries are the main active population. The main difference is in the $\rm 10^{43}\,{<}\,L_{bol}\,{<}\,10^{45}\, erg/s$ population at $z\,{<}\,1$  which is up to 5 times more abundant in the \dm{}. A similar enhancement of sources at low-$z$ is seen for $\rm 10^8\,{<}\,M_{BH,1}\,{<}\,10^9\, M_{\odot}$ with an important increase of equal mass binaries with $\rm 10^{45}\,{<}\,L_{bol}\,{<}\,10^{47}\, erg/s$. 
Finally,  we can see that equal mass systems with $\rm M_{BH,1}\,{>}\,10^9\, M_{\odot}$ take a larger relevance at low-$z$ than in the \fm{}. This is particularly evident at $z\,{\sim}\,1$ for $\rm 10^{45}\,{<}\,L_{bol}\,{<}\,10^{47}\, erg/s$ where the \dm{} finds $0.02$ equal mass binaries per $\rm deg^2$, while the \textit{fiducial} one predicts few $10^{-3}$.\\


The small projected angular separation of our \mbin{} challenges their detection as dual active galactic nuclei. Consequently, without studies addressing periodic lightcurves or Doppler-shifting of AGN broad emission lines, it is not easy to differentiate between AGNs powered by single or binary massive black holes. Taking into account this limitation, we explore the probability that a given observed quasar/AGN is powered by a binary system. To this end, we define the quantity $f_{\rm AGNs}^{\rm Binary}$ as:

\begin{equation}
f^{\rm Binary}_{\rm AGNs}\,{=}\,\rm \frac{N_{MBHBs}(L_{bol}\,{>}\,L^{th}, M_{BH,1})}{N_{AGNs}(L_{bol}\,{>}\,L^{th}, M_{BH,1})}
\end{equation}
where $\rm N_{MBHBs}(L_{bol}\,{>}\,L^{th}, M_{BH,1})$ is the total number of active \mbin{} with primary MBH mass $\rm M_{BH,1}$, and $\rm N_{AGNs}(L_{bol}\,{>}\,L^{th}, M_{BH,1})$ is the total number of accreating \mBHS{} (binaries and single systems) with mass $\rm M_{BH,1}$. Here we assume $\rm L^{th}\,{=}\,10^{43}\, erg/s$ as the threshold to define an active system but we have checked that similar trends are found when we increase $\rm L^{th}$ up to $\rm 10^{45}\, erg/s$. The results presented in Fig.~\ref{fig:FAGNs_Binaries} show that there are only mild differences between \textit{fiducial} and \textit{delayed} models. Regardless of mass, equal mass systems contribute ${<}\,20\%$ to the total population of active \mBHS{}. Conversely, unequal mass binaries have a much larger importance with a contribution that strongly correlates with $\rm M_{BH,1}$. For instance, \mbin{} with $q\,{<}\,0.1$ and $\rm 10^7\,{<}\,M_{BH,1}\,{<}10^8\, M_{\odot}$ can represent at $z\,{=}\,0$ ($z\,{\sim}\,0.5$) up to 40\% (20\%) of active \mBHS{} within their mass range. Larger values are found for binaries with $\rm M_{BH,1}\,{>}10^9\, M_{\odot}$ whose contribution grows up to 60\% as redshift goes to zero. Interestingly, this large importance of unequal mass binaries quickly drops towards high redshift. At $z\,{>}\,1.5$ the value of $f^{\rm Binary}_{\rm AGNs}$ decreases down to $20\%\,{-}\,5\%$, regardless of the mass. \\

\begin{figure}
\centering
\includegraphics[width=1.\columnwidth]{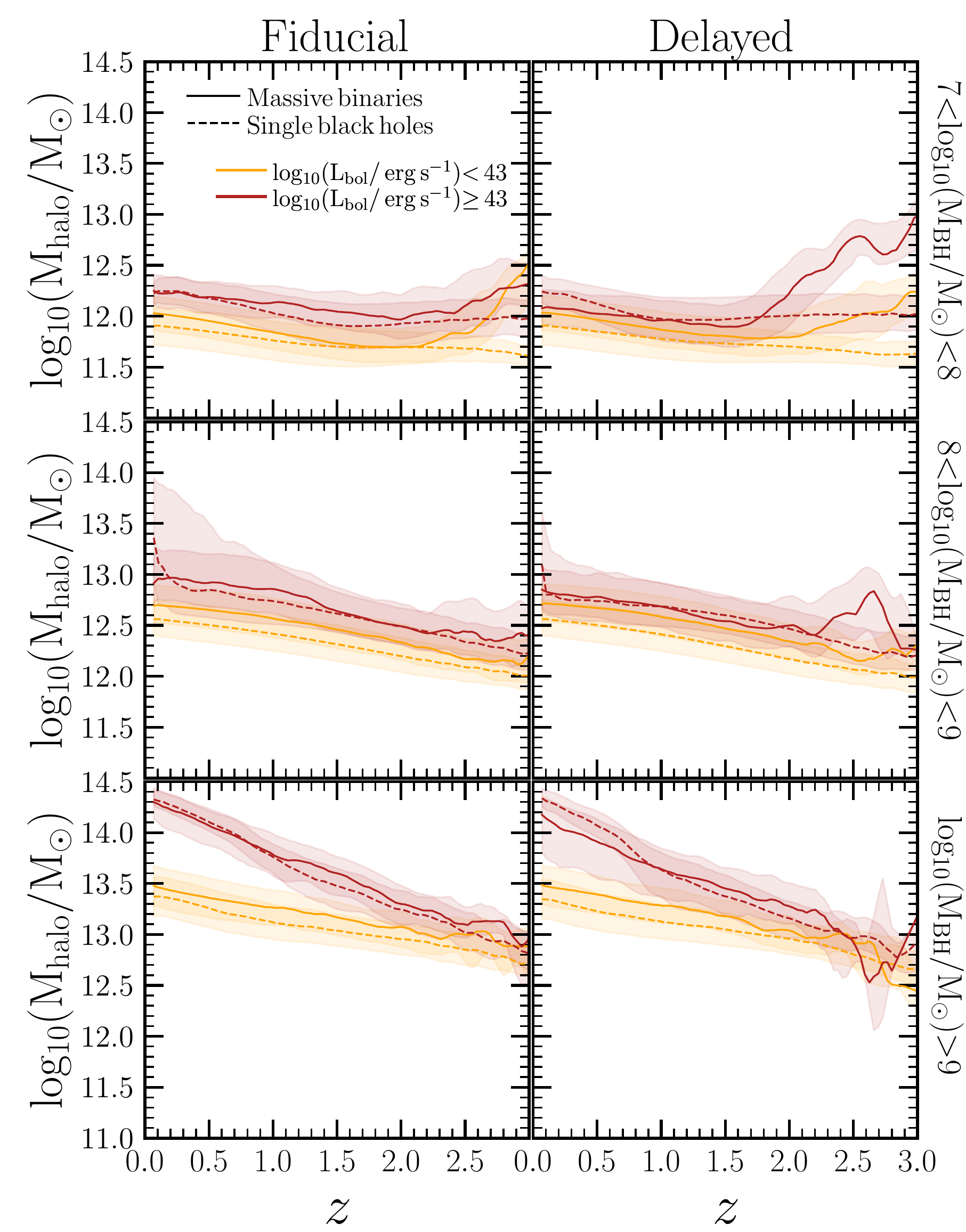}
\caption[]{Median halo mass ($\rm M_{halo}$) of active ($\rm L_{bol}\,{>}\,10^{43} \, erg/s$, red) and inactive ($\rm L_{bol}\,{<}\,10^{43} \, erg/s$, orange) \mbin{} (solid lines) and \mBHS{} (dotted lines). Top, middle and bottom panels represent the results for $\rm 7\,{<}\,log_{10}(M_{BH,1}/M_{\odot})\,{<}\,8$, $\rm 8\,{<}\,log_{10}(M_{BH,1}/M_{\odot})\,{<}\,9$ and $\rm log_{10}(M_{BH,1}/M_{\odot})\,{>}\,9$, respectively. The left panels show the results for the \fm{} whereas the right ones are for the \dm{}. In all the panels the shaded areas correspond to the $\rm 32^{th}\,{-}\,68^{th}$ percentile of the distributions.}
\label{fig:Mhalo_Single_Binaries}
\end{figure}

\begin{figure*}
\centering
\includegraphics[width=1.8\columnwidth]{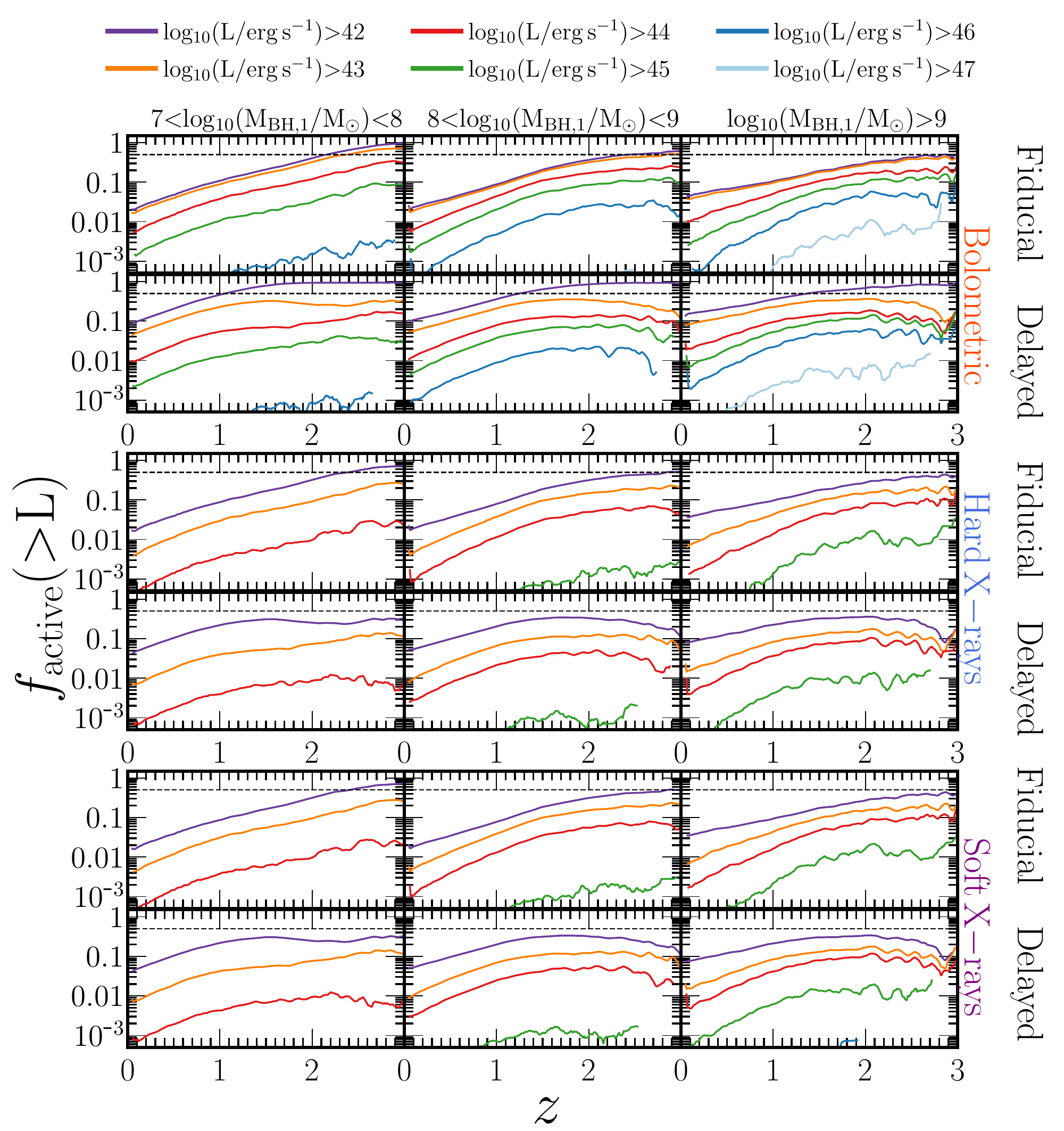}
\caption[]{Redshift evolution of the active fraction of \mbin{}, $f_{\rm active}$, at three different masses of the primary MBH, $\rm M_{BH,1}$: $\rm   7\,{<}\,log_{10}(M_{BH,1}/M_{\odot})\,{<}\,8$ (left panels), $\rm 8\,{<}\,log_{10}(M_{BH,1}/M_{\odot})\,{<}\,9$ (middle panels) and $\rm log_{10}(M_{BH,1}/M_{\odot})\,{>}\,9$ (right panels). Horizontal dotted line highlights the value of $f_{\rm active}\,{=}\,0.5$. Different colors display the predictions when different thresholds in luminosity are assumed. The predictions of $f_{\rm active}$ have been explored for bolometric luminosity (top panels), hard X-rays (middle panels) and soft X-rays (lower panels).}
\label{fig:Factive_MBHB}
\end{figure*}

To explore indirect ways to disentangle between quasars/AGNs powered by \mbin{} and single \mBHS{}, in Fig.~\ref{fig:Mhalo_Single_Binaries} we present the evolution of the hosting halo mass for active ($\rm L_{bol}\,{>}\,10^{43}\, erg/s$) and inactive ($\rm L_{bol}\,{<}\,10^{43}\, erg/s$) \mBHS{} and \mbin{}. Despite the halo mass can be a difficult parameter to estimate from observations, it clearly correlates with the environment, i.e number of galaxies around our target \citep{ValeandOstriker2004,Shankar2006,Conroy2006}. Therefore, the halo mass will give us potential information about the typical overdensities in which these systems are hosted. As shown, regardless of mass and model, the active population of both \mbin{} and single \mBHS{} is systematically hosted in more massive halos than the inactive one. Nevertheless, by comparing single and binary systems we do not see strong differences. Therefore, the results suggest that the environment of quasars/AGNs will not be a good tracer for pointing out the presence of an active \mbin{} with $\rm M_{BH,1}\,{>}\,10^7\, M_{\odot}$. We have checked that the same trends are kept when we divide the population of active \mbin{} between equal and unequal mass.

\subsection{The occurrence of active massive black hole binaries}

In this section we study how frequent is the active phase of \mbin{}. In Fig.~\ref{fig:Factive_MBHB} we present for different $\rm M_{BH,1}$ the fraction of active \mbin{}, $f_{\rm active}$, defined as
\begin{equation}
\rm \mathit{f}_{active} \,{=}\,\frac{N_{MBHBs}(M_{BH,1},{>}\,L_{bol})}{N_{MBHBs}(M_{BH,1})},
\end{equation}
where $\rm N_{MBHBs}(M_{BH,1},{>}\,L_{bol})$ is the number of \mbin{} with $\rm M_{BH,1}$ radiating at luminosity larger than $\rm L_{bol}$. On the other hand, $\rm N_{MBHBs}(M_{BH,1})$ is the total number of \mbin{} with primary MBH mass $\rm M_{BH,1}$. As shown, at $z\,{>}\,2$ the \fm{} predicts that a large fraction of the MBHB population with $\rm 10^7\,{<}\,M_{BH,1}\,{<}\,10^8\, M_{\odot}$ displays some level of gas accretion. For instance, the values of $f_{\rm active}$ can be ${\gtrsim}0.5$ when a threshold of $\rm L_{bol}\,{=}\,10^{43}\, erg/s$ is assumed. However, in the same redshift range, $f_{\rm active}$ can drop down to $0.1$ ($0.01$) when the more strict limiting luminosity $\rm L_{bol}\,{>}\,10^{45}\, erg/s$ ($\rm L_{bol}\,{>}\,10^{46}\, erg/s$) is imposed.  At lower redshifts, the value of $f_{\rm active}$ decreases very quickly at any luminosity threshold, reaching  $f_{\rm active}\,{\leqslant}\,0.1$ by $z\,{<}\,0.5$. This fast drop is caused by the combination of two different processes. The first one is the formation of new \mbin{} embedded in gas poor environments unable to trigger AGN activity. The second is  the fast decrease of gas reservoir around \mbin{} caused by the gas consumption (see Section~\ref{sec:GrowthModels}). This latter effect is particularly evident when we compare the predictions of $f_{\rm active}$ between the \textit{fiducial} and \textit{delayed model}. As shown, the latter one displays values of $f_{\rm active}$ systematically larger regardless of the bolometric luminosity and redshift. When considering $\rm 10^8\,{<}\,M_{BH,1}\,{<}\,10^9\, M_{\odot}$ and $\rm M_{BH,1}\,{>}\,10^9\, M_{\odot}$, we find similar trends to those seen in lighter \mbin{}: at any luminosity there is a decreasing trend of $f_{\rm active}$ towards low-$z$ being earlier and more pronounced in the \fm{}. Interestingly, in both growth models the most luminous \mbin{} ($\rm L_{bol}\,{>}\,10^{45}\, erg/s$) powered by $\rm M_{BH,1}\,{>}\,10^8\, M_{\odot}$  at $z\,{<}\,1$ contribute less than $10\%$ of the total \mbin{} population within their mass range. All these results highlight the challenges of future multi-messenger studies as a consequence of the elusive nature of the most massive \mbin{}.\\

Given current and future facilities exploring the X-ray sky with large sensitivity (see, e.g, eROSITA, \citealt{Merloni2012}, Athena, \citealt{Nandra2013} or Lynx \citealt{Lynx2018}), in Fig.~\ref{fig:Factive_MBHB} we also show the model predictions when using X-rays luminosity as threshold. Specifically, to transform bolometric to soft ($\rm 0.5\,{-}\,2 \,keV$) and hard ($\rm 2\,{-}\,10 \,keV$) X-ray luminosity we make use of the corrections derived in  \cite{Marconi2004}:
\begin{eqnarray} \label{equation:HardSoftXrays}
\rm log_{10}\left(L_{Hx}/L_{bol}\right) \,{=}\, - 1.54 - 0.24\mathcal{L} - 0.012\mathcal{L}^2 + 0.0015\mathcal{L}^3\\
\rm log_{10}\left(L_{Sx}/L_{bol}\right) \,{=}\, - 1.64 - 0.22\mathcal{L} - 0.012\mathcal{L}^2 + 0.0015\mathcal{L}^3
\end{eqnarray}
where $\rm \mathcal{L}\,{=}\,log_{10}(L_{bol}/L_{\sun}) -12$, and $\rm L_{Hx}$ ($\rm L_{Sx}$) is the hard (soft) X-ray luminosity. As we can see, soft and hard X-rays predictions display very similar trends. As for the bolometric luminosity, the predictions differ between the \fm{} and \dm{}, especially at low-$z$. Despite that, both growth models predict that $f_{\rm active}$ rarely overpass values of $0.5$, regardless of redshift and X-ray luminosity.

\begin{figure}
\centering
\includegraphics[width=1.\columnwidth]{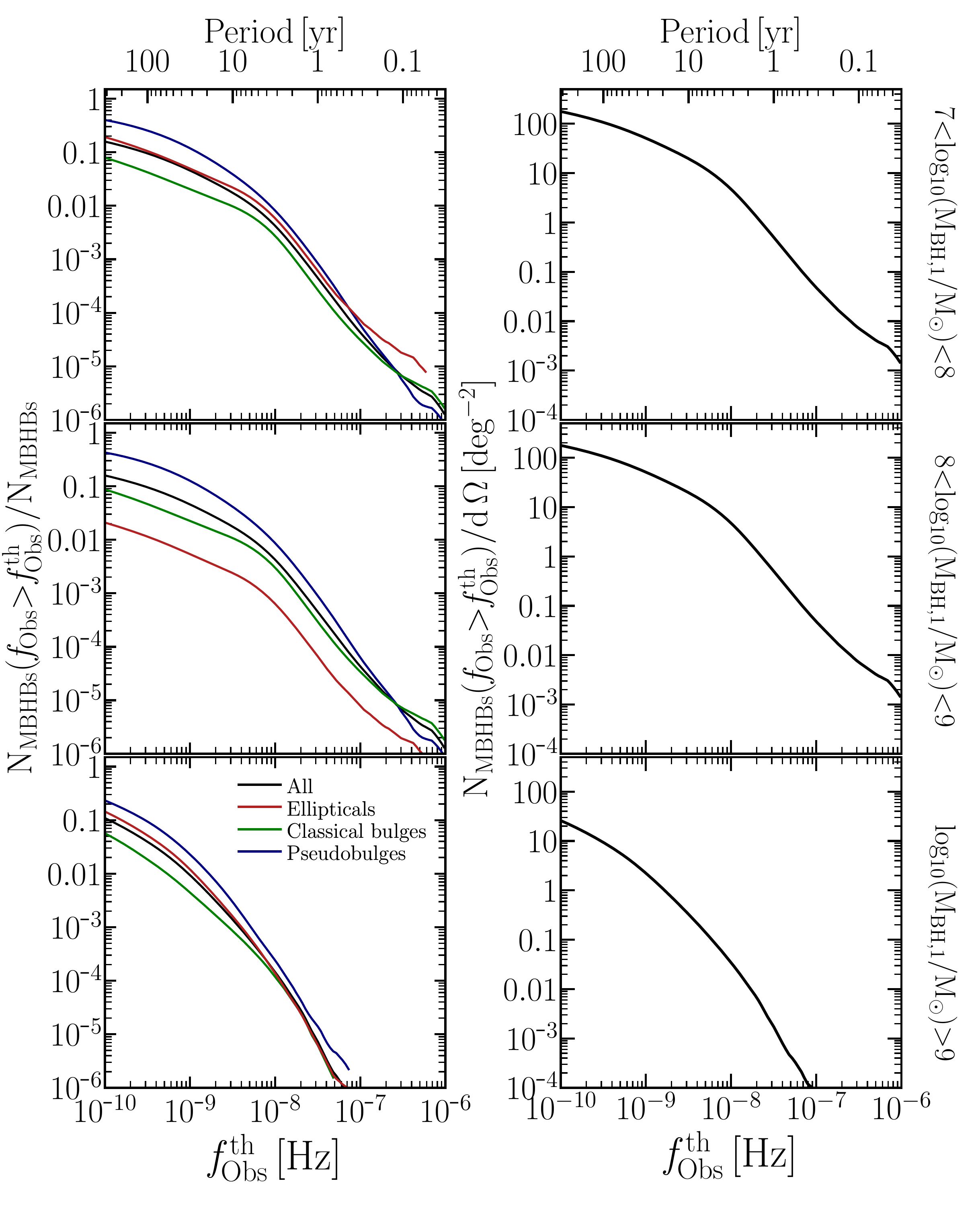}
\caption[]{\textbf{Left panel}: For the \fm{}, the fraction of \mbin{} with observed frequency, $f_{\rm Obs}$, larger than a threshold $f_{\rm Obs}^{\rm th}$. The red, green and blue colors correspond to galaxies with a elliptical-, classical- and pseudo- bulge morphology \textbf{Right panel}: For the \fm{}, the number of binaries per square degree ($\rm deg^2$) emitting GW at a frequency larger than $f_{\rm Obs}^{\rm th}$. Upper, middle and lower panel represent the results for  $\rm 7\,{<}\,log_{10}(M_{BH,1}/M_{\odot})\,{<}\,8$, $\rm 8\,{<}\,log_{10}(M_{BH,1}/M_{\odot})\,{<}\,9$ and $\rm log_{10}(M_{BH,1}/M_{\odot})\,{>}\,9$, respectively.}
\label{fig:N_fobs}
\end{figure}

\subsection{The gravitational wave signature of massive black hole binaries}

Besides having electromagnetic signatures, a large fraction of the \mbin{} studies here are GW sources at nHz frequencies, the regimen probed by PTAs \citep{Sesana2008}. Given that the common red signal recently detected by these experiments might indeed be the stochastic GW background generated by inspiraling supermassive black hole binaries, it is interesting to study what is the frequency distribution of our \mbin{} and which fraction of them will be accessible to PTAs. For that, in the left panels of Fig.~\ref{fig:N_fobs} we present for the \fm{}\footnote{The \dm{} displays the same trends and only minor differences are seen.} the cumulative fraction of \mbin{} emitting GWs above a given observed frequency, $f_{\rm Obs}^{\rm th}$. The specific value of $f_{\rm Obs}$ for each \mbin{} has been computed as:
\begin{equation}
   f_{\rm Obs}\,{=}\,\frac{n}{2\pi (1+z)} \left( (1+\mathit{q}) \frac{\rm M_{BH,1}}{a^3_{\rm BH}}\right)^{1/2}
\end{equation}
where $z$ is the redshift of the binary and $n$ the harmonic number. Despite our model computes self-consistently the evolution of $e_{\rm BH}$, here, for simplicity, we assume that all the \mbin{} are in circular orbits ($e_{\rm BH}\,{=}\,0$) and, thus radiating only at the $n\,{=}\,2$ harmonic. As shown, around 15\% of the \mbin{} are emitting at frequencies $f_{\rm Obs}^{\rm th}\,{>}\,10^{-10}\,\rm Hz$, regardless of $\rm M_{BH,1}$. This percentage drops down to 8\% at $f_{\rm Obs}^{\rm th}\,{>}\,10^{-9}\,\rm Hz$. Interestingly, the specific shape of the distribution depends on the galaxy morphology. About $10\%$ of the MBHB hosted in pseudobulge structures emit GW at frequencies ${>}\,10^{-9}\, \rm Hz$, but this percentage drops up to $2 \, \rm dex$ in the $10^{-9}\,{<}\,f_{\rm Obs}^{\rm th}\,{<}\,10^{-8}\, \rm Hz$ range. For elliptical and classical bulge structures hosting $\rm M_{BH,1}\,{<}\,10^9\, M_{\odot}$, this drop is seen at larger frequencies (${>}\,10^{-8}\, \rm Hz$). However, for more massive systems, elliptical, pseudobulges, and classical bulges display similar behavior. The right panels of Fig.~\ref{fig:N_fobs} show the number of \mbin{} per square degree emitting GWs above a given frequency threshold. For $\rm 10^7 \,{<}\, M_{BH,1}\,{<}\,10^8\, M_{\odot}$ systems, the model predicts up to 100 objects per $\rm \rm deg^2$ with frequencies ${>}\,10^{-9}\, \rm Hz$. Similar values are found for $\rm 10^8 \,{<}\, M_{BH,1}\,{<}\,10^9\, M_{\odot}$. However, binaries with $\rm M_{BH,1}\,{>}\,10^9\, M_{\odot}$ are much rarer, with only ${\sim}\,1$ objects per $\rm \rm deg^2$ at ${>}\,10^{-9}\, \rm Hz$. We stress that an in-depth study on the relation between GW and electromagnetic signatures of the PTA sources is deferred in a companion paper in preparation.

\section{Summary and Conclusions} \label{sec:Conclusions}

In this paper, we studied the hosts and electromagnetic signatures of parsec-scale \mbin{} with primary black hole mass $\rm M_{BH,1}\,{>}\,10^7\, M_{\odot}$. To this purpose, we made use of the \LGalaxies{} semi-analytical model applied on the \texttt{Millennium} merger trees. Specifically, we used the \LGalaxies{} version presented in \cite{IzquierdoVillalba2019LC} and \cite{IzquierdoVillalba2021} where a lightcone construction and several prescriptions for tracing the formation and dynamical evolution of \mbin{} had been included. Since the evolution of the \mBHS{} and \mbin{} depends on the specific prescriptions implemented for describing the mass growth, we explored two different models: a \fm{} where MBHs and \mbin{} exhaust their gas reservoir in a \textit{fast} and efficient way, and a \dm{} allowing for longer periods of gas accretion. These two different models allowed us to explore how the uncertainties in the gas consumption rate affect our predictions. The main results of this work can be summarized as follows:

\begin{itemize}

   \item Number of MBHBs versus redshift: The population of parsec-scale \mbin{} is especially common at $0.5\,{<}\,z\,{<}\,1,$ (${\sim}\,20$ objects per $\rm deg^2$) with unequal mass binaries ($q\,{\leq}\,0.1$) a factor $2\,{-}\,3$ more abundant than equal-mass systems.  In  the \dm{}, as a consequence  of the retarded mass growth, the number of  systems with $q\geq 0.1$ drops by one order of magnitude by $z\sim 3$.\\

   \item  Masses of the host galaxies: Binaries with $\rm M_{BH,1} \,{>}\,10^8\,M_{\odot}$ are located in very massive systems, with halo and stellar masses $\rm {\sim}\,10^{12.5}\, M_{\odot}$ and $\rm {\sim}\,10^{11}\, M_{\odot}$, respectively. Conversely, binaries with $\rm 10^7\,{<}\, M_{BH,1} \,{<}\,10^8 \, M_{\odot}$ are placed galaxies with halos  and stellar masses $\rm {\sim}\,10^{12}\, M_{\odot}$ and $\rm {\sim}\,10^{10}\, M_{\odot}$, respectively. On top of this, single \mBHS{} and \mbin{} populate the same plane of the $\rm M_{BH}\,{-}\,M_{stellar}$ scaling relation, ruling out  that binaries are outliers of the scaling relations.\\

  \item Morphology of the host: Spiral galaxies with a classical bulge component are the preferred hosts of \mbin{} with $\rm M_{BH,1} \,{>}\,10^7\, M_{\odot}$. However, the model predicts that ellipticals become important in the local Universe, especially for systems with $\rm M_{BH,1} \,{>}\,10^9$ where they account for the ${\sim}\,75\%$ of the hosts. Interestingly, \textit{fiducial} and \textit{delayed model} display a compatible behaviour. Thus, the morphology of the galaxy where an MBHB is placed is just driven by the galaxy merger history, regardless of the growth path followed by the binary system.\\  

  \item Occupation fraction: Less than 1\% of galaxies  with $10^9{<}\,\rm M_{stellar}\,{<}\,10^{10}\, M_{\odot}$ host a \mbin{} with $\rm M_{BH,1} \,{>}\,10^7 \, M_{\odot}$. On contrary, up to 10\% of galaxies with $\rm 10^{10}\,{<}\,\rm M_{stellar}\,{<}\,10^{11}\, M_{\odot}$ can host a \mbin{}, with unequal mass binaries of $\rm 10^8\,{<}\,M_{BH,1}\,{<}\,10^9\, M_{\odot}$ the preferred systems. For the most massive galaxies in the model ($\rm M_{stellar}\,{>}\,10^{11}\, M_{\odot}$), the MBHB  occupation fraction can reach up to $50\%$, suggesting that they are the perfect targets for finding low-$z$ parsec-scale binaries. In particular, \mbin{} hosted by these massive galaxies are unequal systems with $\rm M_{BH,1} \,{>}\,10^9 \, M_{\odot}$, but a still significant number of galaxies (${\sim}\,10\%$) host equal mass binaries with $\rm M_{BH,1} \,{>}\,10^8 \, M_{\odot}$. We checked that these results are independent of the growth model explored. \\

  
\item Hardening mechanisms: Regardless of the galaxy morphology, the main shrinking process of binaries with $\rm M_{BH,1}\,{>}\,10^9\, M_{\odot}$ is through encounters with single stars. For lighter systems, the process displays a redshift evolution. While $z\,{<}\,2$ the vast majority of the systems reduce their separation through stellar hardening, at $z\,{>}\,2$ the situation inverts and more than 50\% experience a gas-driven hardening. Interestingly, this percentage is larger in classical bulges than in elliptical and pseudobulge structures.\\

  \item MBHBs in the PTA ($\rm nHz$) band: Looking at the distribution of binary periods, ranging between 100 and 1 years, and their associated GW emission frequency, we found that less than 1\% of \mbin{} hosted in elliptical galaxies and classical bulges have observed frequencies accessible  by PTA experiments ($10^{-9}  \,{<}f_{\rm Obs} \,{<}\,10^{-7}\, \rm Hz$). Moreover, this fraction is systematically smaller in elliptical structures than in classical bulges. Our model indicates that independently of the growth model used, ${\sim}\,90$ binary systems per deg$^2$ should be emitting gravitational waves at the PTA frequencies. \\

  \item MBHBs as active AGNs: In the \fm{} the maximum of the MBHB AGN activity is reached at $z\,{\sim}\,1.5$. In the \dm{}, important accretion events onto \mbin{} are highly suppressed at high redshifts causing a delay of $\rm {\sim}\,2\, Gyr$ in the peak of the activity. In both models, active \mbin{} are mainly unequal-mass systems emitting primarily at sub-Eddington level with luminosity spanning between $\rm 10^{43}\,{<}\,L_{bol}\,{<}\,10^{45}\, erg/s$.  At $z\sim 1$, we find up to $0.2\,{-}\,0.5$ ($1\,{-}\,2$) of these binaries per $\rm deg^2$ in the \textit{fiducial} (\textit{delayed}) \textit{model}. For equal-mass systems, the largest contribution is at $\rm L_{bol}\,{>}\,10^{45}\, erg/s$ but less than $0.06\,{-}\,0.08$ objects per $\rm deg^2$ are predicted.\\

  \item Fraction of active MBHBs: The fraction of active binaries strongly depends on the luminosity threshold. In general, at $z\,{>}\,2$ a large fraction of \mbin{} (${>}\,40\%$) should be in an active phase. On the other hand, at $z<1$ and taking a fiducial threshold of $10^{43}\, \rm erg/s$, less than $10\%$ of the systems can be considered active in any band, including hard and soft X-rays. When comparing the halo mass of the galaxies hosting active \mbin{} and single \mBHS{} we find that both samples display similar values, highlighting that the environments of AGNs are not good tracers to reveal the presence of an active MBHBs.\\
 
\end{itemize}

Given all the results summarized above we can conclude that parsec-scale massive black hole binaries with primary mass ${>}\,10^7\,\msun$ are a common population at $z\,{<}\,1$, being hosted in an important fraction of the most massive galaxies in the local universe. However, most of these \mbin{} are expected to be inactive or very dim, challenging future multi-messenger astronomy studies. Future observations with JWST will be crucial to shed light on that faint AGN population, guiding our understanding of active MBHBs \citep{Seth2021}.  

\section*{Acknowledgements}
We thank the referee for all her/his useful comments which improved the clarity of the paper. The authors warmly thank Silvia Bonoli for the useful discussions and comments. D.I.V. and A.S. acknowledge the financial support provided under the European Union’s H2020 ERC Consolidator Grant ``Binary Massive Black Hole Astrophysics'' (B Massive, Grant Agreement: 818691). D.I.V. acknowledges also financial support from INFN H45J18000450006. M.C. acknowledges funding from MIUR under the Grant No. PRIN 2017-MB8AEZ. This is a pre-copyedited, author-produced PDF of an article accepted for publication in Monthly Notices of the Royal Astronomical Society, following peer review.


\section*{DATA AVAILABILITY}

The simulated data underlying this article will be shared on reasonable request to the corresponding author. This work used the 2015 public version of the Munich model of galaxy formation and evolution: \LGalaxies. The source code and a full description of the model are available at http://galformod.mpa-garching.mpg.de/public/LGalaxies/.


\bibliographystyle{mnras}
\bibliography{references} 





\bsp	
\label{lastpage}
\end{document}